\documentclass{ieeeaccess}
\usepackage{amsmath,amssymb,amsfonts}
\usepackage{algorithmic}
\usepackage{graphicx,dblfloatfix}
\usepackage{textcomp,hyperref}
\usepackage{subcaption}
\usepackage{multirow,makecell}
\usepackage{algorithm}
\hypersetup{urlcolor=blue, colorlinks=true}
\def\BibTeX{{\rm B\kern-.05em{\sc i\kern-.025em b}\kern-.08em
    T\kern-.1667em\lower.7ex\hbox{E}\kern-.125emX}}

\usepackage{pifont}
\newcommand{\cmark}{\ding{51}}%
\newcommand{\xmark}{-}
\newcommand*{\ditto}{---\texttt{"}---}

\usepackage[backend=bibtex,style=ieee,natbib=true,maxcitenames=1,mincitenames=1,uniquelist=false]{biblatex} 
\begin{document}
\history{Date of publication xxxx 00, 0000, date of current version xxxx 00, 0000.}
\doi{xxxx}

\title{CT Perfusion is All We Need: 4D CNN Segmentation of Penumbra and Core in Patients With Suspected Ischemic Stroke}
\author{Luca Tomasetti\authorrefmark{1}, \IEEEmembership{Member, IEEE},
Kjersti Engan\authorrefmark{1}, \IEEEmembership{Senior Member, IEEE}, Liv Jorunn H\o{}llesli\authorrefmark{1,2}, Kathinka D{\ae}hli Kurz\authorrefmark{1,2}, and Mahdieh Khanmohammadi\authorrefmark{1},\IEEEmembership{Member, IEEE}}

\address[1]{Department of Electrical Engineering and Computer Science, University of Stavanger, Stavanger, 4021, Norway}
\address[2]{Stavanger Medical Imaging Laboratory (SMIL), Department of Radiology, Stavanger University Hospital, Stavanger, 4068, Norway}

\tfootnote{All authors are with the BioMedical Data analysis group (\url{https://www.uis.no/en/bmdlab}). This study was supported and approved by the Regional ethic committee project 2012/1499.
}

\markboth
{Tomasetti \headeretal: CTP is All We Need: 4D CNN Segmentation of Penumbra and Core in Patients With Suspected Ischemic Stroke}
{Tomasetti \headeretal: CTP is All We Need: 4D CNN Segmentation of Penumbra and Core in Patients With Suspected Ischemic Stroke}

\corresp{Corresponding author: Luca Tomasetti (e-mail: luca.tomasetti@uis.no).}

\begin{abstract}
Precise and fast prediction methods for ischemic areas comprised of  dead tissue, core, and salvageable tissue, penumbra, in acute ischemic stroke (AIS) patients are of significant clinical interest. They play an essential role in improving diagnosis and treatment planning.
Computed Tomography (CT) scan is one of the primary modalities for early assessment in patients with suspected AIS.
CT Perfusion (CTP) is often used as a primary assessment to determine stroke location, severity, and volume of ischemic lesions.
Current automatic segmentation methods for CTP mostly use already processed 3D parametric maps conventionally used for clinical interpretation by radiologists as input. Alternatively, the raw CTP data is used on a slice-by-slice basis as 2D+time input, where the spatial information over the volume is ignored. 
In addition, these methods are only interested in segmenting core regions, while predicting penumbra can be essential for treatment planning.  
This paper investigates different methods to utilize the entire 4D CTP as input to fully exploit the spatio-temporal information,  leading us to propose a novel 4D convolution layer.
Our comprehensive experiments on a local dataset of 152 patients divided into three groups show that our proposed models generate more precise results than other methods explored.
Adopting the proposed \emph{4D mJ-Net}, a Dice Coefficient of 0.53 and 0.23 is achieved for segmenting penumbra and core areas, respectively.
The code is available on \href{https://github.com/Biomedical-Data-Analysis-Laboratory/4D-mJ-Net}{GitHub}.
\end{abstract}

\begin{keywords}
4D Convolution, Acute Ischemic Stroke, Computed Tomography Perfusion, Deep Neural Network, Image Segmentation.
\end{keywords}

\titlepgskip=-21pt

\maketitle

\section{Introduction}\label{intro}
\PARstart{A}{n} acute ischemic stroke (AIS) generally occurs if a segment of the supplying arteries of the brain is occluded by a blood clot and prevents the regular flow of oxygen-rich blood to the capillaries in the brain tissue.
The ischemic area can roughly be divided into two different types: 1) \emph{penumbra}, areas where the tissue is still vital but critically hypoperfused \citep{astrup1981thresholds}; and 2) \emph{core}, referring to non-salvageable tissue.
If blood flow is not restored timely, penumbra regions may develop rapidly into irreversibly damaged core regions.
Therefore, a fast and accurate understanding of ischemic areas to plan the treatment and tailor further procedures to every single patient is fundamental.

The recommended modalities for diagnostic imaging in AIS patients are Computed Tomography (CT) and Magnetic Resonance Imaging (MRI) \citep{european2008guidelines}.
In the initial stages of an acute ischemic stroke, CT Perfusion (CTP) has proven to be a fast and beneficial tool for evaluating both diagnosis and prognosis \citep{campbell2013ct}.
MRI with Diffusion-weighted imaging (DWI) or non-contrast CT (NCCT) are commonly utilized, after treatment, to assess the \emph{final infarct areas} (FIAs) \citep{bivard2014defining}.   
These imaging modalities are obtained hours or days after the patient’s treatment.

CTP is a four-dimensional (4D) spatio-temporal examination to assess the passage of blood in the brain.
It is performed by acquiring a series of three-dimensional (3D) CT scans of a specific portion of the brain at time intervals during contrast agent injection.
By using an iodinated contrast agent, density changes in the brain tissue over time can be analyzed. 
The shape and height of the time density curve depend on the brain tissue's perfusion \citep{kurz2016radiological}.
The raw 4D CTP contains vast data; hence, detecting ischemic strokes can be challenging for neuroradiologists, primarily due to the need for precise diagnosis and for rapid treatment decisions.  

To overcome this challenge, medical doctors rely on a set of clinically interpretable parametric maps (PMs) generated from the 4D CTP scan \citep{kurz2016radiological, kudo2010differences}.
The generated PMs reduce the temporal dimension and produce 3D volumes.
Commonly used PMs are cerebral blood flow (CBF), cerebral blood volume (CBV), time-to-maximum (TMAX), and time-to-peak (TTP). 
CBF represents the blood supply in the brain at a given time;
CBV refers to the blood volume present at a given time in a brain region;
TMAX is the flow-scaled residue function in the tissue; while TTP shows the time until the contrast agent reaches the tissue \citep{kurz2016radiological}.
Maximum intensity projection (MIP) is also usually generated. 
MIP images are calculated as the maximum Hounsfield unit (HU) value over the time sequence of the CTP, providing a 3D volume from the 4D acquisition of CTP.
Although PMs provide helpful information about ischemic brain tissue, extracting them from the 4D CTP scans limits the spatio-temporal information only to specific subsets of information \citep{soltanpour2022using}.
Several methods \citep{d2015time,bivard2014defining,lin2016whole,mcverry2014derivation,cereda2016benchmarking} have used thresholding techniques to predict the ischemic areas from the PMs.
However, simple thresholding approaches over-simplify the complexity in AIS \citep{nielsen2018prediction,bivard2022does}.

In the past years, Deep Neural Networks (DNNs), and especially Convolutional Neural Networks (CNNs), have been successfully applied in numerous medical applications: image classification tasks \citep{chen2021annotation, neel2022quantifying, fuster2022invasive, kanwal2022devil, kanwal2023}, automatic video analysis \citep{chakraborty2013video, meinich2020activity}, and activity recognition \citep{akbulut2020wearable, ahmed2023transforming, amft2008recognition}.
Automatic image segmentation adopting U-Net structure \citep{ronneberger2015u} and its numerous variants have produced innovative outcomes for several applications \citep{isensee2021nnu, zhou2018unet++, wetteland2020multiscale, khan2022rms, kamnitsas2017efficient}.

Several DNNs have been proposed for AIS applications to predict and segment \textit{only} the FIAs using CT studies in combination with PMs derived from CTP scans as input \citep{clerigues2019acute, abulnaga2018ischemic, rava2021investigation, soltanpour2019ischemic, soltanpour2021improvement}. 
Other researchers have proposed architectures to segment the ischemic lesion (i.e., the core) from the images obtained at hospital admission.
\citet{kasasbeh2019artificial} were the first to implement a CNN with a set of PMs as input for ischemic core segmentation.
\citet{tomasetti2023self} proposed a few-shot self-supervised architecture for hypoperfused (core + penumbra) tissue segmentation using a combination of PMs and raw scans as input of the model. The work demonstrated the feasibility of using self-supervised techniques for the segmentation of this type of tissue.
\citet{werdiger2023machine} introduced a machine learning segmentation method to delineate hypoperfused tissue, demonstrating the capabilities of this methodology over classic thresholding approaches. They used four PMs as input features for their model.
However, a general problem with all the methods mentioned above is relying on commercial CTP software and using heavily pre-processed information (i.e., PMs) rather than taking advantage of the totality of the raw 4D CTP scans.

DNNs are more suitable for discovering information from raw data.
Nevertheless, relying on raw data (directly exploiting the temporal and spatial dimensions) is scarcely explored in the literature for AIS applications. The task is challenging because of the low contrast and low signal-to-noise ratio in the CTP scans. 
Relatively few studies proposed DNN models with encouraging results, exploiting the temporal dimension to assess acute stroke lesions using 4D CTP scans.
\citet{soltanpour2022using} utilized CTP images to create 2D matrices in which each row is a voxel, and each column is a time point. 
The 2D matrices are used as input for a model that shows encouraging results in differentiating healthy tissue from FIAs.
\citet{de2023perfu} promoted a 2D+time symmetry-aware CNN-based architecture to segment FIAs using solely CTP scans. Their work estimated the irreversibly damaged areas, demonstrating the possibilities of using 4D CTP images for this task. 
\citet{bertels2019contra} used a U-Net-like structure for segmenting FIAs using CTP scans as an input plus contra-lateral information. Results were promising, but further research is needed due to their far-from-ideal registration of the contra-lateral information.
\citet{de2020differentiable} introduced a two-step model for estimating FIAs using the 4D CTP series as input. 
They first generate an arterial input function and later deconvolve it with a singular value decomposition approach to find the infarction.
\citet{amador2021stroke} designed a framework based on Temporal Convolution Network to predict AIS FIAs from 4D CTP studies. Due to memory constraints, they independently processed each 2D slice of the 4D CTP dataset. In their recent work, \citet{amador2022predicting} also proposed an extension of their model where 3D+time tensors of the ipsilateral stroke hemisphere are used as input to predict FIAs.
\citet{robben2020prediction} proposed a DNN that predicts the FIAs directly from 4D raw CTP plus patients metadata.
Their proposed architecture relied on a series of 3D Convolution layers; the input is a list of 4D CTP scans sampled at different resolutions.  
Their method presented promising segmentation results; however, their main target was to estimate final infarct volume allowing clinicians to simulate different treatments and gain insight for the procedures. They were not taking into consideration the penumbra in their study. Plus, the quality of the ground truth images is debatable since they rely on NCCT follow-up images acquired between 24 hours and five days after patient's admission.
It has been reported that FIAs can grow after 24 hours in NCCT measurements \citep{bucker2017associations}.

All of the segmentation methods mentioned above rely on ground truth labels obtained from DWI and/or NCCT hours or days after the patient's admission since they are predicting FIAs.
Even though follow-up images (DWI and NCCT) represent the gold standard for estimating core \citep{bivard2014defining}, there are some limitations with these techniques \citep{schellinger2010evidence,goyal2020challenging}.
Follow-up images can only be used to assess FIAs but not penumbra regions.
Plus, some studies have demonstrated that the detected FIAs can be partially reverse in DWI performed in an early time window  \citep{goyal2020challenging,kidwell2000thrombolytic,labeyrie2012diffusion}.

The studies of \citep{soltanpour2022using, de2023perfu, bertels2019contra, robben2020prediction, de2020differentiable, amador2021stroke, amador2022predicting} indicate the potential of 4D data in AIS prediction. 
However, they all consider FIAs and an appropriate method is still required to simultaneously handle the spatio-temporal information for segmenting the ischemic \emph{core} and \emph{penumbra} regions. 
Understanding the penumbra's extension during the ischemic stroke's first stages is crucial for treatment decision \citep{murphy2006identification,tomasetti2021machine}.
To the best of our knowledge, our work in Tomasetti \textit{et al.} \citep{tomasetti2021machine} using machine learning, and later in Tomasetti \textit{et al.} \citep{tomasetti2020cnn,tomasetti2022multi} using DNN, were the first and only to segment \textit{both} core and penumbra areas. In \citep{tomasetti2021machine,tomasetti2022multi} the PMs were used as input, in \citep{tomasetti2020cnn} 2D + time CTP images were segmented slice-by-slice. 

In this paper, we present and investigate three novel models to segment the two ischemic regions (core and penumbra), where the input is the entire 4D CTP scans arranged in different ways to exploit the spatio-temporal nature of the data. 
We compare all models with previous work based on PMs \citep{tomasetti2022multi} and slice-by-slice CTP \citep{tomasetti2020cnn}, and two methods proposed by \citet{amador2021stroke, amador2022predicting}.


The main contributions of this work can be summarized in four points:
\begin{enumerate}
\item We develop a novel 4D convolution layer that processes the 4D raw data. 
\item We implement the 4D convolution layer and propose a DNN model, \emph{4D mJ-Net}, to segment ischemic core and penumbra areas from 4D CTP scans.
\item We extend previous methods \citep{amador2021stroke,tomasetti2020cnn} to be used directly with 4D CTP scans as input.
\item To assess the results, we use manual annotations obtained by two expert neuroradiologists from the 4D CTP data upon patients' admission. We also demonstrate the feasibility of our proposed methods by comparing their performances with existing models that rely on different inputs.
\end{enumerate}

\section{Data Material}\label{material}

A section of the brain is repeatedly scanned during the passage of 40 ml iodine-containing contrast agent (Omnipaque 350 mg/ml) and 40 ml isotonic saline in a cubital vein with a flow rate of 6 ml/s to highlights changes in the tissue; the scan delay was four seconds.
Each brain slice contains a fixed number of time points $t_{\text{max}}$ representing the temporal dimension.
The width and height of each image are $512\times512$ pixels with a resolution of 0.4258 mm/pixel and a slice thickness of 5mm.
The first twenty time points are acquired every 1s, while the remaining ten images are every 2s.

CTP scans from 152 patients collected between January 2014 and August 2020 formed the dataset.
137 of these patients had an AIS with a visible perfusion deficit.
During the diagnostic workup, the remaining 15 patients were admitted with suspected strokes but were determined not to have suffered from a stroke episode after the diagnostic workup.
Raw perfusion data from the CTP examination was used to generate PMs with the software ``syngo.via'' from Siemens Healthineers, with manufacturer default settings. 
The arterial input function was automatically selected, with few exceptions where it was chosen manually (e.g., severe cardiac failure).  

The patients were divided into the following groups: 77 patients with large vessel occlusion (LVO), 60 patients with non-large vessel occlusion (Non-LVO), and the remaining 15 patients without ischemic stroke (WIS).
Based on CT angiography, LVO was defined as occlusion of any of the following arteries: the internal carotid artery, M1 and proximal M2 segment of the middle cerebral artery, A1 segment of the anterior cerebral artery, P1 segment of the posterior cerebral artery, basilar artery, and vertebral artery occlusion.
Non-LVO was defined as patients with perfusion deficit with more distal artery occlusion or with perfusion deficit without visible artery occlusion.
\begin{table}[ht]
\caption{Division in training, validation, and test dataset.}
\label{tab:division}
\centering
\resizebox{\columnwidth}{!}{
\begin{tabular}{l|c|c|c|c}
\Xhline{3\arrayrulewidth}
& \textbf{Training (\#; \%)} & \textbf{Validation (\#; \%)} & \textbf{Test (\#; \%)} & \textbf{Tot. (\#; \%)} \\
\Xhline{3\arrayrulewidth}
\textbf{LVO} & 42; 54.5 & 16; 20.8 & 19; 24.7 & 77; 50.6 \\
\hline
\textbf{Non-LVO} & 36; 60 & 13; 21.7 & 11; 18.3 & 60; 30.5 \\
\hline
\textbf{WIS} & 9; 60 & 3; 20 & 3; 20 & 15; 9.8 \\
\Xhline{3\arrayrulewidth}
\textbf{Total} & 87; 58.6 & 32; 19.7 & 33; 21.7 & 152; 100 \\
\hline
\end{tabular}
}
\end{table}
The dataset is randomly split into a training, validation, and test set.
The percentage of the three subsets (LVO, Non-LVO, WIS) is equally distributed among the sets, as shown in Table \ref{tab:division}.

\subsection{Ground truth}\label{gt}
The manual annotations are based on the entire CT dataset, including the PMs derived from CTP. MRI performed during the first days after hospital admission was also utilized.
Two expert neuroradiologists manually annotated ground truth images by utilizing the complete set of the CT examination (NCCT, CT angiography, and CTP), which includes PMs from the CTP (CBV, CBF, TTP, TMAX) and the MIP images. 
The PMs were visually assessed. 
In general, ischemic regions with increased TTP and TMAX and reduced CBF but preserved CBV were considered as penumbra, while areas with additionally reduced CBV were deemed as core. 
Additionally, the MRI examination, including DWI, was conducted within 1 to 3 days after the CT examination, and clinical information was used to assist in generating the ground truth images. 
The annotations were performed using an in-house developed software in Matlab\footnote{The code is publicly available at \url{https://github.com/Biomedical-Data-Analysis-Laboratory/CTP-Matlab}}.


\section{Background theory}

\subsection{Notation}
Table \ref{tab:notation} presents the various formal notations adopted in the remainder of the paper.
Let the data obtained from a CTP scan be defined as a 4D tensor $V \in \mathbb{R}^{(X \times Y \times Z \times T)}$.
After a series of pre-processing steps (details in Sec. \ref{sec:preproc}), we define the 4D tensor as $\widetilde{V} \in \mathbb{R}^{(X \times Y \times Z \times T)}$.
The four dimensions of a CTP scan are defined as width ($X$), height ($Y$), depth ($Z$), and time ($T$).
The list of time points in the time dimension is given by $t = [t_j | \forall j \in \{1,2,\cdots, t_{\text{max}}\}]$, where $t_{\text{max}}$ is the last time point of the list.
We indicate how the notation superscript adopts the time dimension in the various inputs.
Furthermore, we define $z = [z_i | \forall i \in \{1,2,\cdots, z_{\text{max}}\}]$ as the list of brain slices in the depth dimension, where $z_{\text{max}}$ corresponds to the last slice.
We illustrate how the depth dimension is being used in the inputs through the notation subscript.
Fig. \ref{fig:inputs} displays the input combination of all the techniques.

\begin{table}[ht]
\caption{List of formal notations used in the paper.}
\label{tab:notation}
\centering
\resizebox{\linewidth}{!}{
\begin{tabular}{c|c}
\Xhline{3\arrayrulewidth}
\textbf{Notation} & \textbf{Description} \\
\Xhline{3\arrayrulewidth}
$t = [t_j | \forall j \in \{1,2,\cdots, t_{\text{max}}\}]$ & List of all the time points. \\ \hline
$t_{\text{max}}$ & \makecell{ Last time point in the \\ time dimension. } \\ \hline
$z = [z_i | \forall i \in \{1,2,\cdots, z_{\text{max}}\}]$ & List of all the slices. \\ \hline
$z_{\text{max}}$ & \makecell{Last slice in the \\ depth dimension.} \\ \hline
$\mathcal{I} = \{i-1,i,i+1\}$ & \makecell{Set of indexes $i$, plus its \\ neighbours $i-1$ and $i+1$.} \\ \hline
$z_I = \{z_{i-1},z_i,z_{i+1}\}$ & \makecell{Set of slices: \\ $ z_{i-1},z_i,z_{i+1} $.} \\ \hline
$V \in \mathbb{R}^{(X \times Y \times Z \times T)} $ & 4D \textbf{raw} CTP input. \\ \hline
$\mathcal{C} = \{$healthy brain, penumbra, core$\}$ & Set of classes. \\ \hline
$I^{t_j}_{z_i} \in \mathbb{R}^{(X \times Y)}$ & \makecell{2D brain slice $z_i$ \\ at time point $t_j$.} \\ \hline
$P_{z_i} \in \mathbb{R}^{(X \times Y)}$ & \makecell{2D probability output \\ of brain slice $z_i$.} \\ \hline
$\widetilde{\cdot}$ & \makecell{Input \textbf{after}\\ \textbf{pre-processing steps}.} \\ \hline
$\bar{\cdot}$ & \makecell{\textbf{List} of inputs} \\ \hline
$\varphi(\cdot)$ & Concatenation function. \\ \hline 
$\widehat{\cdot} $ & \makecell{\textbf{Concatenated} inputs after \\ passing through $\varphi(\cdot)$.} \\ \hline
$\bar{V}^{t}_{z_i} = [ \widetilde{I}_{z_i}^{t_j} | \forall t_j \in t ] \in \mathbb{R}^{(X \times Y)}$ & \makecell{List of 2D images $\widetilde{I}_{z_i}$ \\ for all the time points $t$. \\ Input for \emph{2D-TCN} (Sec. \ref{tcn}).} \\ \hline
$\widehat{V}_{z_i}^t =\varphi(\widetilde{I}_{z_i}^{t_j} | \forall t_j \in t) \in \mathbb{R}^{(X \times Y \times T)}$ & \makecell{2D+time volume of slice $z_i$. \\ Input for \emph{mJ-Net} (Sec. \ref{25dmj}).} \\ \hline
$\widehat{V}^{t_j}_{z_\mathcal{I}} = \varphi(\widetilde{V}^{t_j}_{z_{i-1}},\widetilde{V}^{t_j}_{z_i},\widetilde{V}^{t_j}_{z_{i+1}}) \in \mathbb{R}^{(X \times Y \times Z)}$ & \makecell{3D volume of slices $z_\mathcal{I}$ \\ for a time point $t_j$.} \\ \hline
$\bar{V}^{t}_{z_\mathcal{I}} = [\widehat{V}^{t_j}_{z_\mathcal{I}} | \forall t_j \in t] \in \mathbb{R}^{(X \times Y \times Z)}$ &\makecell{ List of 3D volumes of slices $z_\mathcal{I}$ \\ for all the time points $t$. \\ Input for \emph{3D-TCN} (Sec. \ref{3dtcn}).} \\ \hline
$\bar{V}_{z_\mathcal{I}}^t = [\widehat{V}_{z_{i-1}}^t,\widehat{V}_{z_i}^t,\widehat{V}_{z_{i+1}}^t] \in \mathbb{R}^{(X \times Y \times Z \times T)}$ & \makecell{ List of 2D+time volumes \\ for slices $z_\mathcal{I}$. Input \\ for \emph{3D+time mJ-Net} (Sec. \ref{3dmj}). } \\ \hline
$\widehat{V}^t_{z_\mathcal{I}} = \varphi(\widehat{V}_{z_{i-1}}^t,\widehat{V}_{z_i}^t,\widehat{V}_{z_{i+1}}^t) \in \mathbb{R}^{(X \times Y \times Z \times T)}$ & \makecell{4D Tensor of slices $z_\mathcal{I}$ \\ over all the time points $t$. \\ Input for \emph{4D mJ-Net} (Sec. \ref{4dmj}).} \\ \hline
\end{tabular}
}
\end{table}

All methods return a 3D output, segmenting the images $P_{z_i}$ slice-by-slice.
The segmented 2D image $P_{z_i}$ corresponds to a brain slice $z_i$ at index $i$.
The predicted image $P_{z_i}$ contains brain tissue segmented with the classes $\mathcal{C}$ (if any): healthy brain, penumbra, and core.

\subsection{Pre-processing steps}\label{sec:preproc}

The 4D CTP dataset underwent a series of pre-processing steps to extract brain tissue from the raw CTP scans.
The raw CTP studies are saved as DICOM files.
The pre-processing steps 
can be summarized as follow: 
\begin{enumerate}
    \item Co-registration of all the images in the 4D CTP scan using the first time point image as the frame of reference in order to correct possible motion artifacts. An intensity-based image registration with similarity transformation was used in this step.
    \item All the registered CTP scans were encoded into HU values to have a known quantitative scale to describe radiodensity efficiently.
    To calculate the HU value for a voxel $V$ with a rescale slope (RS) and a rescale intercept (RI) extracted from the DICOM header: $V(x,y,z,t)_{\text{HU}} = V(x,y,z,t) \cdot \text{RS} + \text{RI}$  
    \item Brain extraction of CT studies plays an essential role in stroke imaging research \citep{smith2002fast,najm2019automated}. An automatic brain extraction method designed by \citet{najm2019automated} was selected for this purpose due to its proven efficiency with CT datasets and public availability.
    \item Gamma correction ($\gamma=0.5$) and histogram equalization were also performed after step 3.
    \item Finally, z-score ($z$) on the enhanced 4D tensor is applied to normally distribute the data. 
\end{enumerate}

The input for all the methods (except for the \emph{Multi-input PMs}'s approach) follows the same pre-processing steps.
These steps were performed to improve the quality of the images by enhancing the contrast.
An additional resampling step for all the images was performed to ensure uniform distribution in temporal dimension.
The effects of the pre-processing steps and re-sampling are examined in an ablation study in Sec. \ref{sec:ablation}.

\subsection{Convolution in many dimensions}\label{4dconv}

A handful of DNN methods have been proposed to exploit 4D data with a full 4D Convolution (4D-Conv) layer.
\citet{gessert2020deep} and \citet{bengs20204d} proposed a 4D spatio-temporal convolutional network to optical coherence tomography force estimation. They demonstrated that using the full 4D data information yields better performances than 3D data. 
\citet{myronenko20194d} introduced a 4D CNN to segment cardiac volumetric sequences using CT scans, showing the advantages of using their proposed architecture compared to a classic 3D CNN.    

In the remainder of this manuscript, let define $I(x,y,z,t) \in \mathbb{R}^4, \mathcal{H}(w,h,d,p) \in \mathbb{R}^4$ as a 4D tensor and a 4D kernel, respectively. 
The $x$ and $w$ indicate the width of the 4D structures; $y$ and $h$ express the height dimension; $z$ and $d$ define the depth dimension, while $t$ and $p$ represent the time dimension of the 4D structures.
Like a 3D Convolution (3D-Conv) can be represented as the sum of multiple 2D Convolution (2D-Conv) along the depth dimension, a 4D-Conv operation can be described as the sum of multiple 3D-Conv along the temporal dimension.
The loop rearrangement to avoid repeated 3D-Conv operations allows a true (non-separable) 4D convolution operation \citep{myronenko20194d}.

Let define a 2D-Conv $g''(x,y)$, where a 2D input image $I(x,y) \in \mathbb{R}^2$ is convolved with a 2D kernel $\mathcal{H}(w,h) \in \mathbb{R}^2$ as:
\begin{align*}
g''(x,y) &= \mathcal{H}(w,h) \circledast I(x,y) \\
&= \sum_{i=0}^{w-1} \sum_{j=0}^{h-1} \mathcal{H}(i,j) I(x+\widetilde{w}-i,y+\widetilde{h}-j)
\end{align*}
where $\circledast$ is the convolution operation.
$\widetilde{h} \equiv \lfloor \frac{1}{2}(h-1) \rfloor,\widetilde{w} \equiv \lfloor \frac{1}{2}(w-1) \rfloor$ correspond to the half-width, and half-height of the kernel $\mathcal{H}$.

\begin{figure}[h!]
\begin{minipage}[b]{.8\columnwidth}
\centerline{\includegraphics[width=\columnwidth]{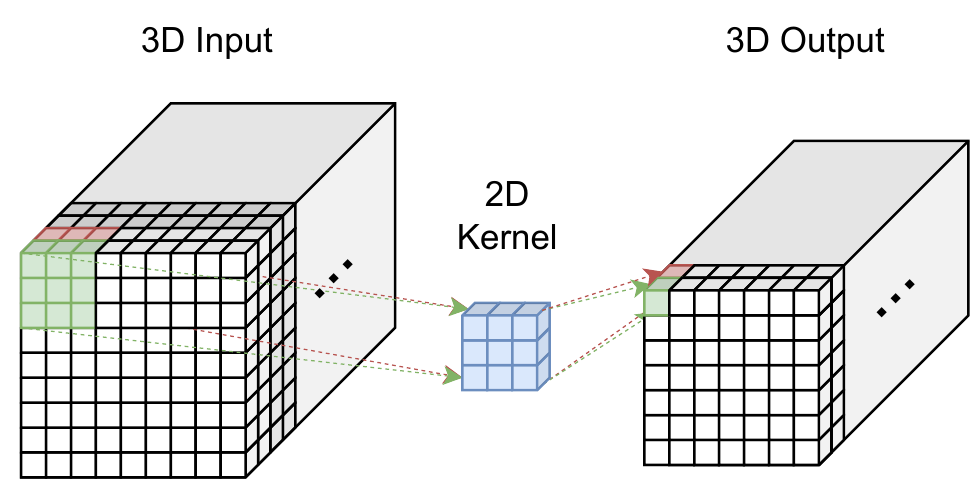}}
\subcaption{2D-Conv with a 3D input and a 2D kernel}
\label{fig:convex:2d}
\end{minipage}
\begin{minipage}[b]{.8\columnwidth}
\centerline{\includegraphics[width=\columnwidth]{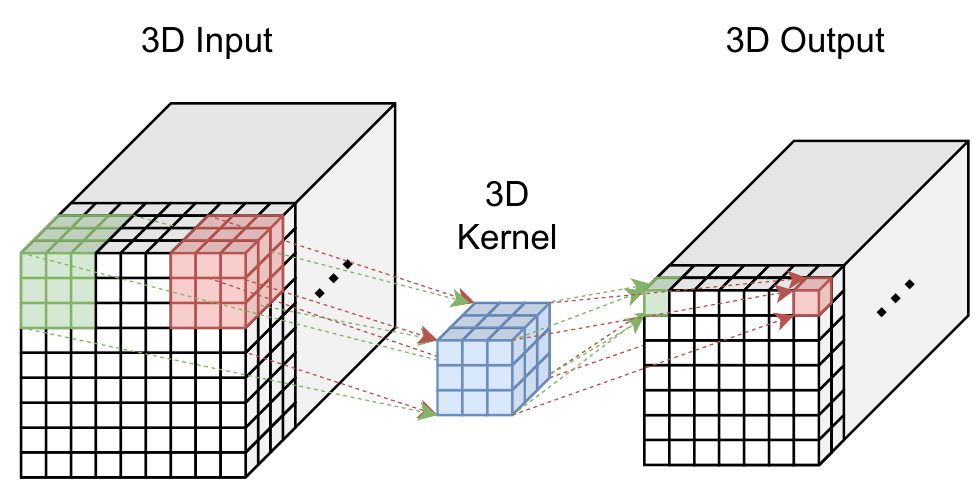}}
\subcaption{3D-Conv with a 3D input and a 3D kernel}
\label{fig:convex:3d}
\end{minipage}
\caption{Visual examples of (a) 2D-Conv and (b) 3D-Conv. Both examples utilize a 3D input, and no padding is applied.
Each box corresponds to a pixel value.}
\label{fig:convex}
\end{figure}

Thereafter, let us define a 2D-Conv $g''(x,y,z)$ with a 2D kernel $\mathcal{H}(w,h) \in \mathbb{R}^2$ and a 3D input $I(x,y,z) \in \mathbb{R}^3$. Since the convolution operation is performed slice by slice over the third dimension, the 3D input can be seen as a list of 2D input $I(x,y,z) = \{ I(x,y,z_m) | \forall m \in \{1, \dots, z_\text{max}\} \}$, where $I(x,y,z_m) \in \mathbb{R}^2$ are the coordinates $(x,y)$ at slice $z_m$:
\begin{align*}
g''(x,y,z) &= \mathcal{H}(w,h) \circledast I(x,y,z)  \\
g''(x,y,z_m) &= \sum_{i=0}^{w-1} \sum_{j=0}^{h-1} \mathcal{H}(i,j) I(\tilde{x},\tilde{y}, z_m) \; \forall m \in \{1, \dots, z_\text{max}\}
\end{align*}
where $ \tilde{x} \equiv x+\widetilde{w}-i$ and $\tilde{y} = y+\widetilde{h}-j$.
Fig. \ref{fig:convex:2d} presents a visual representation of the 2D-Conv $g''(x,y,z)$ with a 3D input and a 2D kernel.


Furthermore, let define a 3D-Conv operation $g'''(x,y,z)$ of a 3D kernel $\mathcal{H}(w,h,d) \in \mathbb{R}^3$ and a 3D input volume $I(x,y,z) \in \mathbb{R}^3$ as:
\begin{align*}
g'''(x,y,z) = \mathcal{H}(w,h,d) \circledast I(x,y,z) \\
= \sum_{k=0}^{d-1} \sum_{i=0}^{w-1}\sum_{j=0}^{h-1} \mathcal{H}(i,j,k) I(x+\widetilde{w}-i, y+\widetilde{h}-j, z+\widetilde{d}-k)
\end{align*}
where $\widetilde{d} \equiv \lfloor \frac{1}{2}(d-1) \rfloor$ is the half-depth of $\mathcal{H}$. Fig. \ref{fig:convex:3d} displays an example of a 3D-Conv with a 3D input and 3D kernel. 
If we define the 3D input $I(x,y,z)$ as before, then the output of a 3D-Conv $g'''(x,y,z)$ can be rewritten as the sum of multiple 2D-Conv operations over the third dimension:
\begin{align*}
g'''(x,y,z) &= \mathcal{H}(w,h,d) \circledast I(x,y,z) \\
&= \sum_{k=0}^{d-1} \mathcal{H}(w,h,k)  I(x,y,z+\widetilde{d}-k) \\
&= \sum_{k=0}^{d-1} g''(x,y,z+\widetilde{d}-k)
\end{align*}
where $\mathcal{H}(w,h,k) \in \mathbb{R}^2$ is a 2D kernel at index $k$, while $I(x,y,z+\widetilde{d}-k) \in \mathbb{R}^2$ is a 2D image at slice $z+\widetilde{d}-k$ where $[z+\widetilde{d}-k \in z_m | \forall m \in \{1, \dots, z_\text{max}\}]$.
The output dimension is defined as: $\text{dim}(g''') = \text{dim}(I) - \text{dim}(\mathcal{H}) + 1$; thus, if the $\text{dim}(I) \equiv \text{dim}(\mathcal{H})$, then $\text{dim}(g''') = 1$.

Moreover, lets define a 3D-Conv operation $g'''(x,y,z,t)$ of a 3D kernel $\mathcal{H}(w,h,d) \in \mathbb{R}^3$ with a 4D input tensor $I(x,y,z,t) \in \mathbb{R}^4$. The 4D tensor $I(x,y,z,t)$ can be seen as a list of $I(x,y,z,t) = [ I(x,y,z,t_n)  | \forall n \in \{1, \dots, t_\text{max}\} ]$, where each element in the list corresponds to the coordinates $(x,y,z)$ of a $t_n$ element in the temporal dimension. 
Then, the 3D-Conv operation $g'''(x,y,z,t)$ can be seen as: 
\begin{align*}
g'''(x,y,z,t) = \mathcal{H}(w,h,d) \circledast I(x,y,z,t) \\
= \sum_{k=0}^{d-1} \sum_{i=0}^{h-1} \sum_{j=0}^{w-1} \mathcal{H}(i,j,k) I(\tilde{x},\tilde{y}, \tilde{z}, t_n) \; \forall n \in \{1, \dots, t_\text{max}\}
\end{align*}
where $ \tilde{x} \equiv x+\widetilde{w}-i$, $\tilde{y} = y+\widetilde{h}-j$, and $\tilde{z} \equiv z+\widetilde{d}-k$.

If we use a 2D+time kernel $\mathcal{H}(w,h,p) \in \mathbb{R}^3$ and we define $I(x,y,z,t)$ as a list of $I(x,y,z,t) = [ I(x,y,z_m,t) | \forall m \in \{1, \dots, z_\text{max}\} ]$, where each element in the list corresponds to the coordinates $(x,y,t)$ of a slice $z_m$, then the 3D-Conv operation $g'''(x,y,z,t)$ can be rewritten as:
\begin{align*}
g'''(x,y,z,t) = \mathcal{H}(w,h,p) \circledast I(x,y,z,t) \\
g'''(x,y,z_m,t) = \mathcal{H}(w,h,p) \circledast I(x,y,z_m,t) \; \forall m \in \{1, \dots, z_\text{max}\} \\
= \sum_{l=0}^{p-1} \mathcal{H}(w,h,l) I(x,y,z_m,t+\widetilde{p}-l) \; \forall m \in \{1, \dots, z_\text{max}\}
\end{align*}
where $\widetilde{p} \equiv \lfloor \frac{1}{2}(p-1) \rfloor$ is the temporal dimension of the kernel $\mathcal{H}$ halved.

A 4D-Conv $g''''(x,y,z,t)$ of a 4D input $I(x,y,z,t) \in \mathbb{R}^4$ and a 4D kernel $\mathcal{H}(w,h,d,p) \in \mathbb{R}^4$ can be defined as:
\begin{align*}
g''''(x,y,z,t) = \mathcal{H}(w,h,d,p) \circledast I(x,y,z,t) \\
= \sum_{l=0}^{p-1} \sum_{k=0}^{d-1} \sum_{i=0}^{w-1}\sum_{j=0}^{h-1} \mathcal{H}(i,j,k,l) I(\tilde{x},\tilde{y},\tilde{z},\tilde{t})
\end{align*}
 where $ \tilde{x} \equiv x+\widetilde{w}-i$, $\tilde{y} = y+\widetilde{h}-j$, $\tilde{z} \equiv z+\widetilde{d}-k$, and $\tilde{t} \equiv t+\widetilde{p}-l$.

\begin{figure}[h!]
\centering
\centerline{\includegraphics[width=\linewidth]{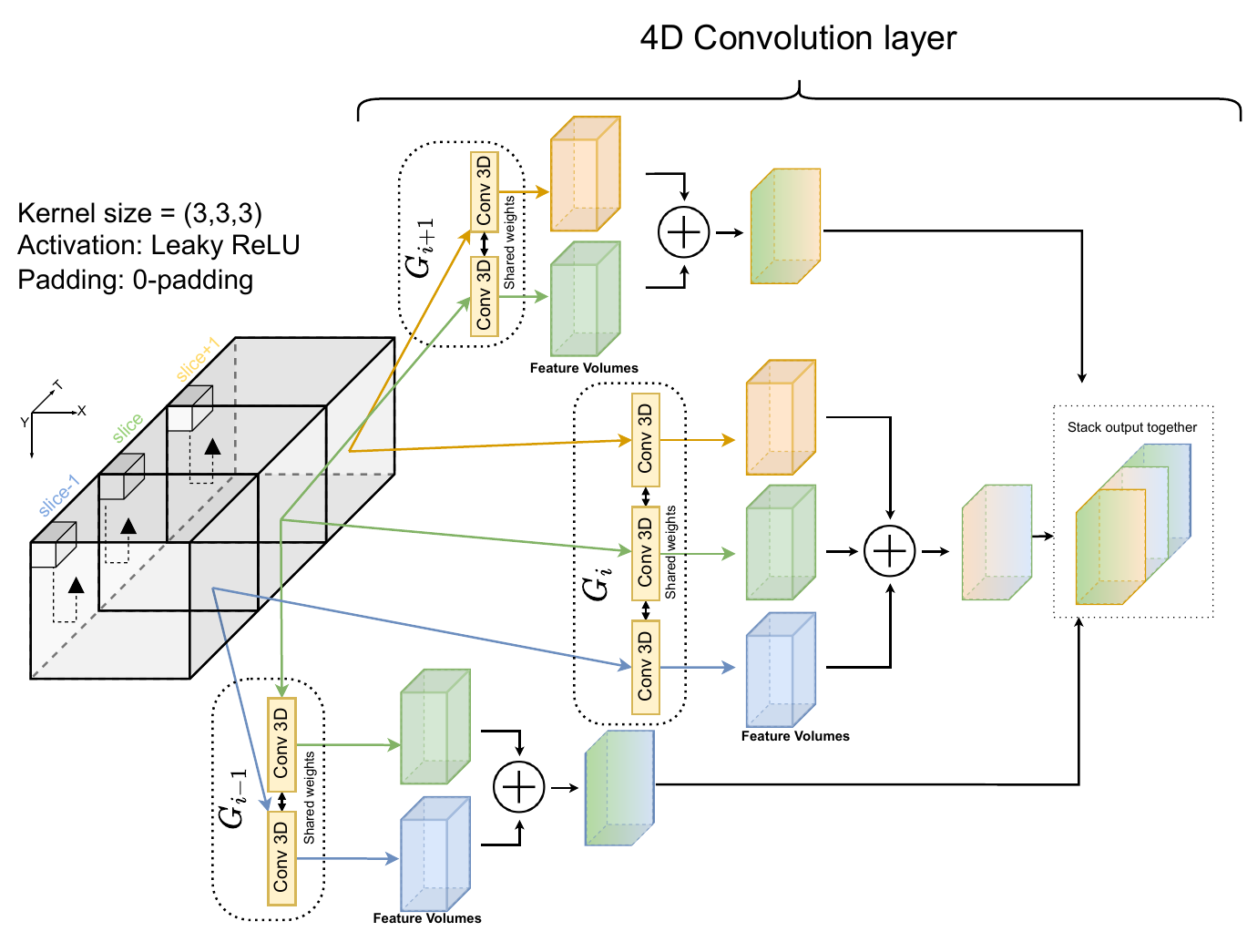}}
\caption{Visual representation of a 4D-Conv layer.
The input is $\widehat{V}^t_{z_\mathcal{I}} = \varphi(\widehat{V}_{z_{i-1}}^t,\widehat{V}_{z_i}^t,\widehat{V}_{z_{i+1}}^t) \in \mathbb{R}^{(X \times Y \times Z \times T)}$.
A series of 3D-Conv operations are calculated over a 4D input.
Several groups $(G_{i-1}, G_i, G_{i+1})$ are generated, one for each volume involved in the operation.
}
\label{fig:4Ddetails}
\end{figure}

Finally, in a similar way, we define a 4D-Conv $g''''(x,y,z,t)$ and a 4D kernel $\mathcal{H}(w,h,d,p)$ as the sum of multiple 3D-Conv over a specific dimension, i.e., the third dimension. 
The 4D kernel $\mathcal{H}(w,h,d,p)$ can be seen as a list of $\mathcal{H}(w,h,k,p) | \forall k \in d-1$, where $\mathcal{H}(w,h,k,p)$ is a 2D+time volume of the $k$th slice over the entire $p$ elements in the temporal dimension:
\begin{align*}
g''''(x,y,z,t) &= H(w,h,d,p) \circledast I(x,y,z,t) \\
&= \sum_{k=0}^{d-1}  \mathcal{H}(w,h,k,p) I(x,y,z+\widetilde{d}-k,t) \\
&= \sum_{k=0}^{d-1} g'''(x,y,z+\widetilde{d}-k,t) \\
\end{align*}
where $I(x,y,z+\widetilde{d}-k,t) \in \mathbb{R}^3$ is a 2D+time volume at slice $z+\widetilde{d}-k$ over the totality $t$ element in the temporal dimension, and $[z+\widetilde{d}-k \in z_m | \forall m \in \{1, \dots, z_\text{max}\}]$

Fig. \ref{fig:4Ddetails} gives a visual explanation of the proposed 4D-Conv layer.
The input for our 4D-Conv layer is a 4D tensor $\widehat{V}^t_{z_\mathcal{I}}$: the 2D+time volume of the $i$th brain slice $\widehat{V}^t_{z_i}$ over all the time points $t$, and the two 2D+time volumes of the neighboring brain slices ($\widehat{V}^t_{z_{i-1}}$ and $\widehat{V}^t_{z_{i+1}}$).
The 4D-Conv layer uses a loop rearrangement with three 3D-Conv groups $(G_{i-1}, G_i, G_{i+1})$, one for each volume slice involved in the operation.
In each group $G_i$, several 3D-Conv layers are created.
All convolution layers in each group shared the weights.
Each 3D-Conv layer is used for a single input volume.
The number of layers depends on the legal subscript indexes, i.e., for the group $G_{i-1}$, there are two 3D-Conv layers since the legal subscript indexes are $\{i-1,i\}$: the indexes are given by the current volume slice $z_{i-1}$ and the only neighboring volume slice $z_i$.
The output of each group is a set of feature volumes summed together.
The resulting feature volumes are stacked together to keep the same dimension as the input.

\section{Existing methods \& Proposed 4D methods}
In this paper, we present three novel deep learning (DL) approaches that accommodate 4D input data.
In the remainder of the paper, they are called \emph{3D-TCN} (Sec. \ref{3dtcn}), \emph{3D+time mJ-Net} (Sec. \ref{3dmj}), and \emph{4D mJ-Net} (Sec. \ref{4dmj}).

\begin{figure*}[h!]
\centering
\centerline{\includegraphics[width=.84\linewidth]{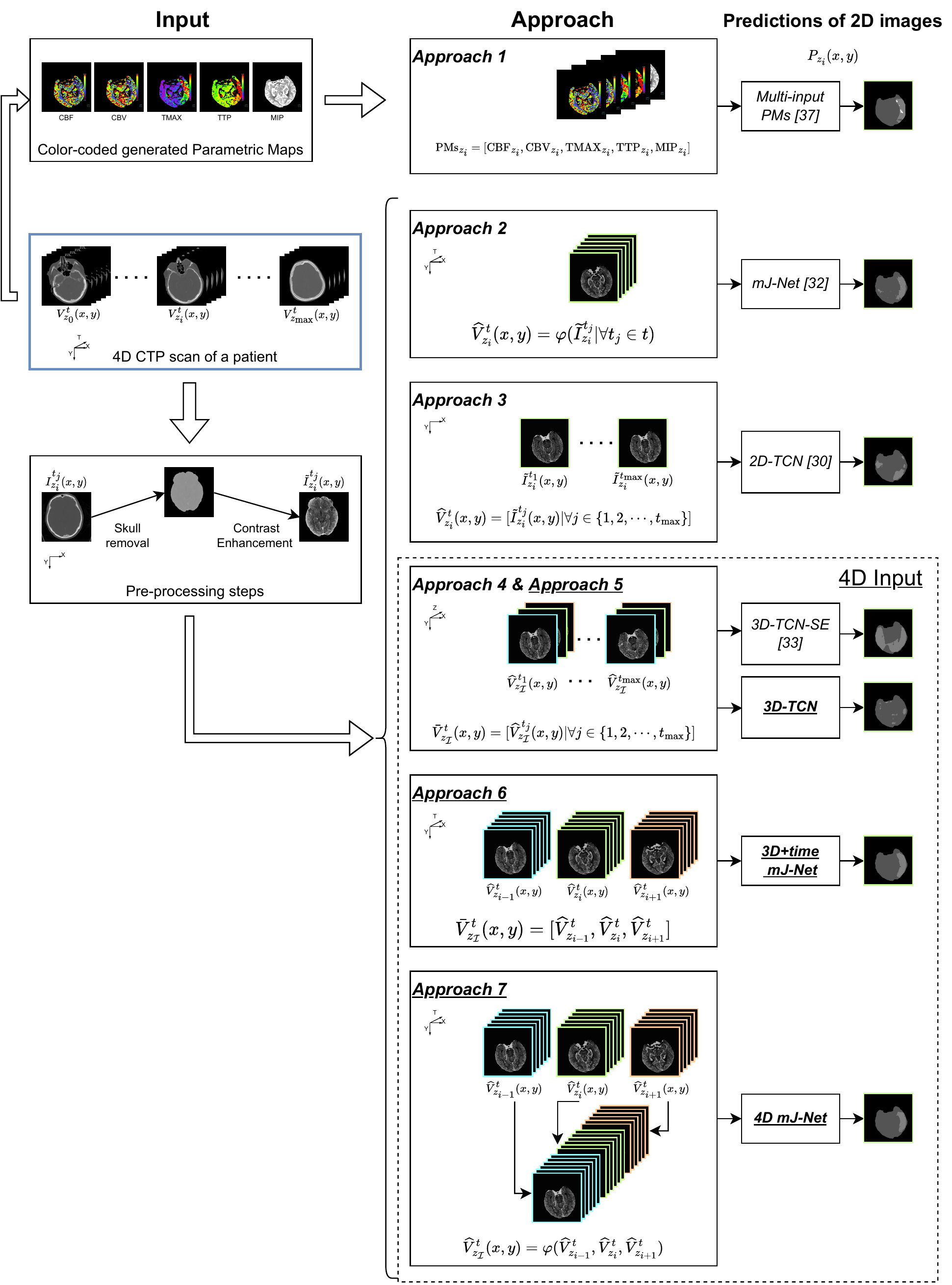}}
\caption{Visual comparison of the input for each implemented approach.
Every 4D CTP patient's study $V \in \mathbb{R}^{(X \times Y \times Z \times T)}$ undergoes a series of pre-processing steps to enhance each CTP scan. 
Approach 1 (\emph{Multi-input PMs}) \citep{tomasetti2022multi} accepts a list of PMs generated from a CTP study in input. 
Approach 2 (\emph{mJ-Net}) \citep{tomasetti2020cnn} use a 2D+time volume $\widehat{V}_{z_i}^t (x,y)$ as input.
Approach 3 (\emph{2D-TCN}) and Approach 4 (\emph{3D-TCN-SE}) follows the model proposed by \citet{amador2021stroke} and \citet{amador2022predicting}, respectively.
The resulting 4D tensor is fed to one of the approaches.
The proposed approach 5 (\emph{3D-TCN}), approach 6 (\emph{3D+time mJ-Net}), and approach 7 (\emph{4D mJ-Net}) take in input the entire 4D CTP processed data $\widetilde{V}$.
}
\label{fig:inputs}
\end{figure*}

Together with the proposed approaches, we implemented and compared the \emph{2D-TCN} \citep{amador2021stroke} (Sec. \ref{tcn}), the \emph{3D-TCN-SE} \citep{amador2022predicting} (Sec. \ref{sec:3dtcnse}), and the \emph{mJ-Net} \citep{tomasetti2020cnn} (Sec. \ref{25dmj}) to validate the performances of our models.
We also compare the models with a method that uses a set of PMs as input \citep{tomasetti2022multi}.
In the remainder of the paper, we call this architecture \emph{Multi-input PMs} (Sec. \ref{multipms}).
Fig. \ref{fig:inputs} visually compares the input utilized for the various approaches. 

\subsection{Existing methods}\label{sec:exmet}

\subsubsection{Approach 1: Multi-input PMs}\label{multipms}
The \emph{Multi-input PMs} model was proposed in \citep{tomasetti2022multi}. 
This architecture was used as a baseline study because all the PMs were input for the model.
The input for the architecture is a list of PMs for each brain slice $z_i$: $\text{PMs}_{z_i}$, as shown in Table \ref{tab:methods}.
The loss function implemented is the Focal Tversky loss (FTL) \citep{abraham2019novel}; for a specific class $c$, the FTL is defined as:
\begin{equation*}
\text{FTL}(x,y) = \sum_c^{\mathcal{C}} (1 - \text{TI}_c)^{1/\gamma}
\end{equation*}
where $\gamma \ge 1$ is a hyper-parameter that forces the loss function to focus more on less accurate predictions that have been misclassified \cite{abraham2019novel}.
Denoting $x_{i,c} \in [0,1]$ as the probability of the $i$th predicted pixel to belong to class $c$; $y_{i,c} \in \{0,1\}$ as the pixel $i$ with class $c$ in a ground truth image, $\text{TI}_c$ is the Tversky index (TI) for a class $c$ defined as:
\begin{equation*}
\text{TI}_c = \frac{\sum_{i=1}^{M\times N} x_{i,c}y_{i,c}}{\sum_{i=1}^{M\times N} x_{i,c}y_{i,c} + \alpha \sum_{i=1}^{M\times N} \widehat{x_{i,c}}y_{i,c} + \beta \sum_{i=1}^{M\times N} x_{i,c}\widehat{y_{i,c}}}
\end{equation*}
where $\widehat{x_{i,c}} = 1-x_{i,c}$ is the probability that the $i$th pixel is not of class $c$, and $\widehat{y_{i,c}} = 1-y_{i,c}$ represents the complement of pixel $i$ in a ground truth image.
The hyper-parameters $\alpha$ and $\beta$ control the trade-off between precision and recall.
We refer the reader to \citep{tomasetti2022multi} for a more extensive explanation and discussion about this approach.

\subsubsection{Approach 2: mJ-Net}\label{25dmj}
The \emph{mJ-Net} approach was proposed in \citep{tomasetti2020cnn}.
As presented in Table \ref{tab:methods}, the input $\widehat{V}_{z_i}^t (x,y)$ for \emph{mJ-Net} is a 2D+time volume of the same brain slice $z_i$ at index $i$ over all the time points $t$.
We define the dimension of this input as 2D+time; the first dimension of the input is time.

The loss function used for this method is the soft Dice Coefficient loss (SDCL) \citep{milletari2016v}.
The SDCL is a modified version of the Dice Coefficient score mainly used in medical domains where the classes to predict are highly unbalanced due to a small region of interest compared to the background of the scans.
The SDCL can be written as:
\begin{align*}
\text{SDCL}(x,y) = \sum_c^{\mathcal{C}} \left(1 - \frac{2 \sum_i^{M\times N} x_{i,c} y_{i,c}}{\sum_i^{M\times N} x_{i,c}^2 + \sum_i^{M\times N} y_{i,c}^2}\right)
\end{align*}
The first section of the model contains 3D-Conv layers to extract information from the temporal dimension, while the second part follows the classic U-Net structure \citep{ronneberger2015u}.
For more details about the \emph{mJ-Net} approach, we refer the reader to \citep{tomasetti2020cnn}.

\subsubsection{Approach 3: 2D Temporal Convolutional Network}\label{tcn}

\begin{figure}[h!]
\centering
\begin{minipage}[b]{\columnwidth}
\centering
\centerline{\includegraphics[width=\columnwidth]{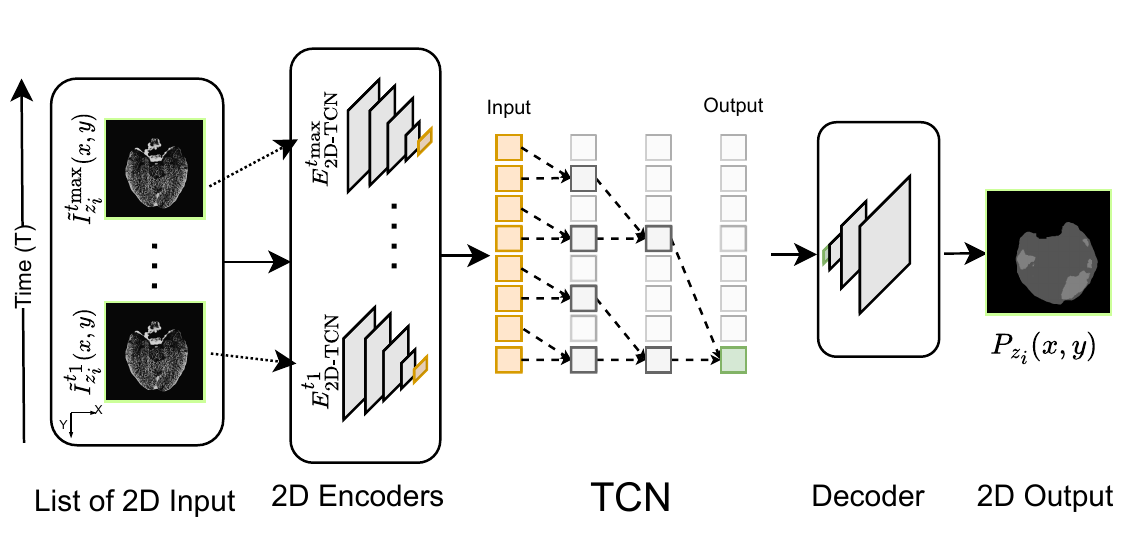}}
\subcaption{\emph{2D-TCN} architecture proposed by \citet{amador2021stroke}.}
\label{fig:tcn:2d}
\end{minipage}
\begin{minipage}[b]{\columnwidth}
\centering
\centerline{\includegraphics[width=\columnwidth]{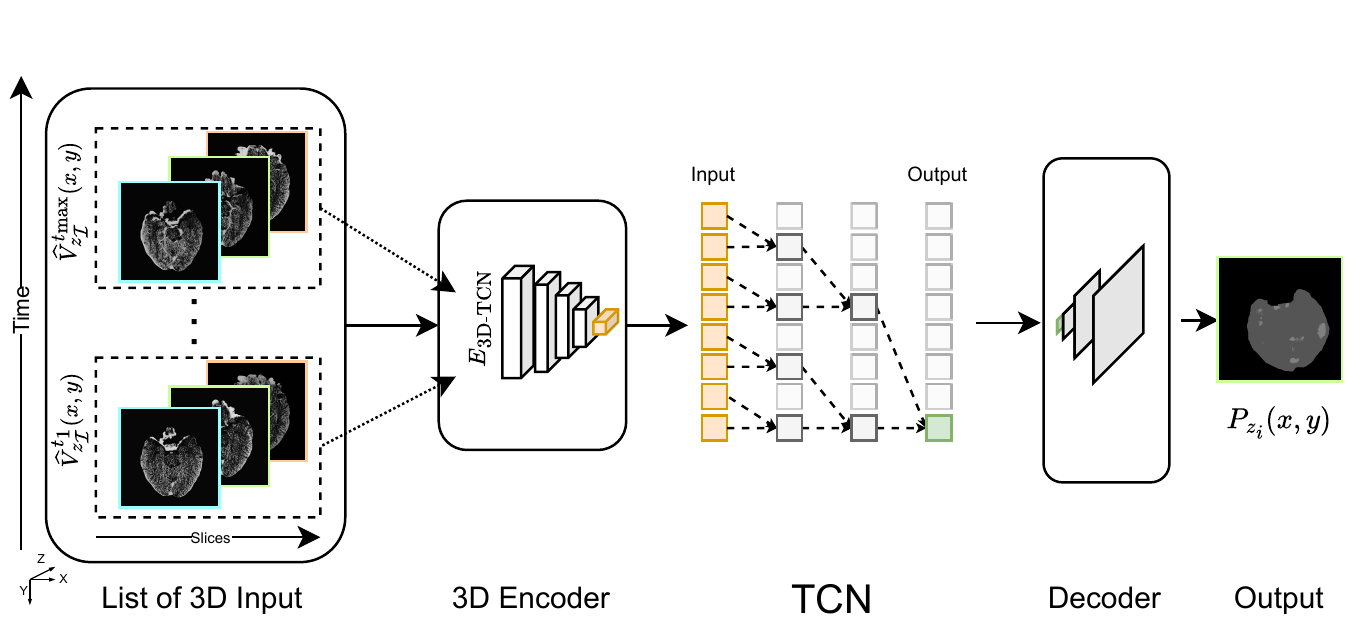}}
\subcaption{\emph{3D-TCN-SE} proposed by \citet{amador2022predicting}}
\label{fig:tcn:3d}
\end{minipage}
\caption{Visual comparison between (a) the \emph{2D-TCN} architecture \citep{amador2021stroke} and (b) the \emph{3D-TCN-SE} \citep{amador2022predicting}.}
\label{fig:tcn}
\end{figure}

For comparison reasons, we implemented the method proposed by \citet{amador2021stroke}.
We call this architecture \emph{2D-TCN} in the remainder of the paper.
Fig. \ref{fig:tcn:2d} provides a general architecture overview.
They proposed a Temporal Convolutional Network (TCN), which has been shown to outperform conventional neural networks in different tasks \citep{bai2018empirical}.
Moreover, a TCN has a lower memory requirement for training than other Recurrent Neural Networks \citep{bai2018empirical}.

The \emph{2D-TCN} was trained with the exact implementation as the original work (see Table \ref{tab:methods} for more details). 
The \emph{2D-TCN} model receives the 4D CTP scans in input re-sampled to 1 second per time point.
The 4D input is processed as a list of 2D brain slices $z_i$ for each time point $t$.
Thus, the actual input for the \emph{2D-TCN} is a list $\bar{V}^{t}_{z_i}$, as mentioned in Table \ref{tab:methods}. 
The list $\bar{V}^{t}_{z_i}$ contains all the time points of the brain slice $z_i$.
Every 2D input image of the list $ \tilde{I}_{z_i}^{t_j}$ at time point $t_j$ is fed to a 2D encoder $E^{t_j}_{\text{2D-TCN}}$ to extract features in the latent space.
Each $E^{t_j}_{\text{2D-TCN}}$ encoder returns a ($4\times4\times \text{Ch}$) feature vector, where $\text{Ch}$ corresponds to the number of channels.
The architecture merges the low-level feature vectors across the different $t_j$ time points to capture the spatio-temporal information.
The merged feature vector $\text{ETOT}_{\text{2D-TCN}} = [E^{t_j}_{\text{2D-TCN}} | \forall t_j \in t]$ is used as input to the TCN, which yields a one-dimensional vector $O_{\text{2D-TCN}}$ of 64 elements.
Finally, a decoder takes the $O_{\text{2D-TCN}}$ and generates a final 2D image $P_{z_i}(x,y)$.
The Dice Coefficient loss (DCL), the same as the original paper, was implemented as the loss function as follows: 
$$
\text{DCL}(x,y) = \sum_c^\mathcal{C} \left(1 - \frac{2 \sum_i^{M\times N} x_{i,c} y_{i,c}}{\sum_i^{M\times N} x_{i,c} + \sum_i^{M\times N} y_{i,c}}\right)
$$
For more details about the \emph{2D-TCN} approach, we refer the reader to \citep{amador2021stroke}.

\subsubsection{Approach 4: 3D Temporal Convolutional Network Single Encoder}\label{sec:3dtcnse}
We implemented a similar method from \citet{amador2022predicting}. We call this approach \emph{3D-TCN-SE} due to using a single encoder (SE).
Fig. \ref{fig:tcn:3d} shows a simplified version of the proposed architecture, emphasizing the input difference between the \emph{2D-TCN} and this model.

The \emph{3D-TCN-SE} model receives the 4D CTP scans re-sampled to 1 second per time point. 
The input is a list of $t$ 3D volumes $\bar{V}^{t}_{z_\mathcal{I}}$ (Table \ref{tab:methods}). 
Each 3D input volume in the list $\widehat{V}^{t_j}_{z_\mathcal{I}}(x,y)$ corresponds to the concatenation of the $i$th brain slice $z_i$ plus its neighbouring slices $z_{i-1}$ and $z_{i+1}$ over a specific time point $t_j$.
The \emph{3D-TCN-SE} approach uses a single encoder $E_{\text{3D-TCN}}$ for all the elements in the input list. 
It is worth mentioning that the \emph{3D-TCN-SE} model is trained with the entire brain images for comparison reasons and not with just the ipsilateral hemisphere, as in the original paper \citep{amador2022predicting}.

\begin{figure}[h!]
\centering
\begin{minipage}[b]{.9\columnwidth}
\centering
\centerline{\includegraphics[width=\columnwidth]{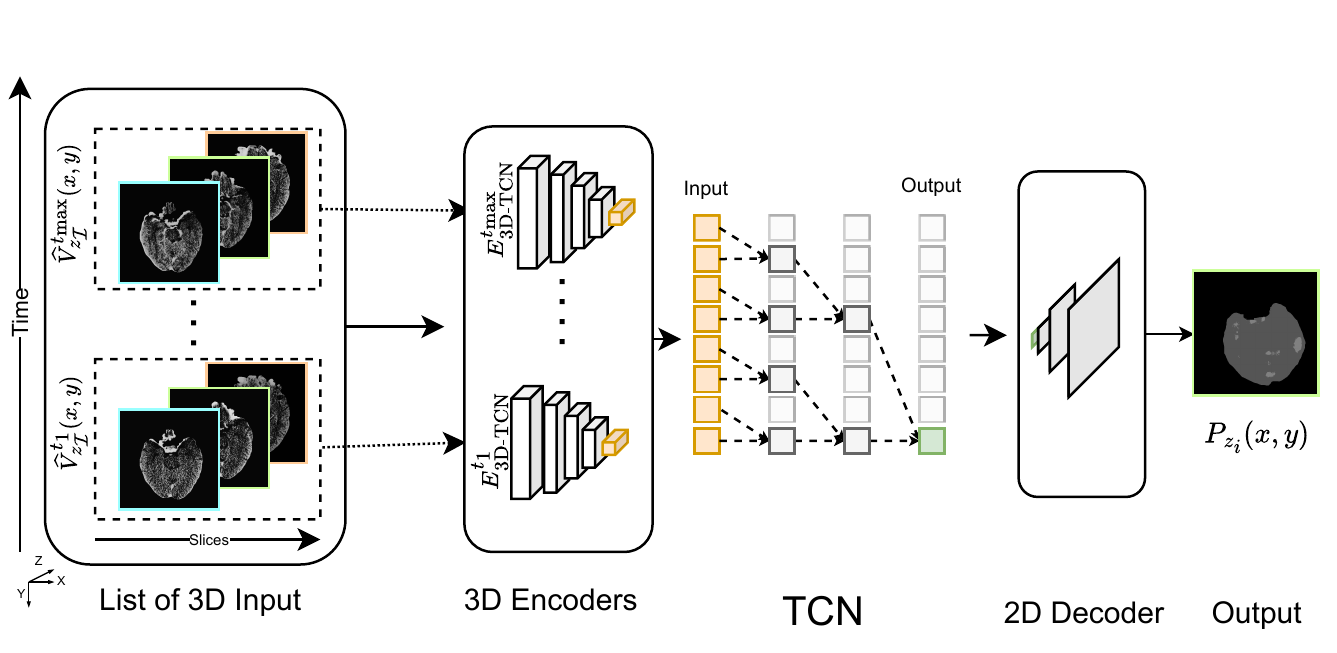}}
\subcaption{\emph{3D-TCN} architecture (Sec. \ref{3dtcn})}
\label{fig:comparison:tcn}
\end{minipage}
\begin{minipage}[b]{.9\columnwidth}
\centering
\centerline{\includegraphics[width=\columnwidth]{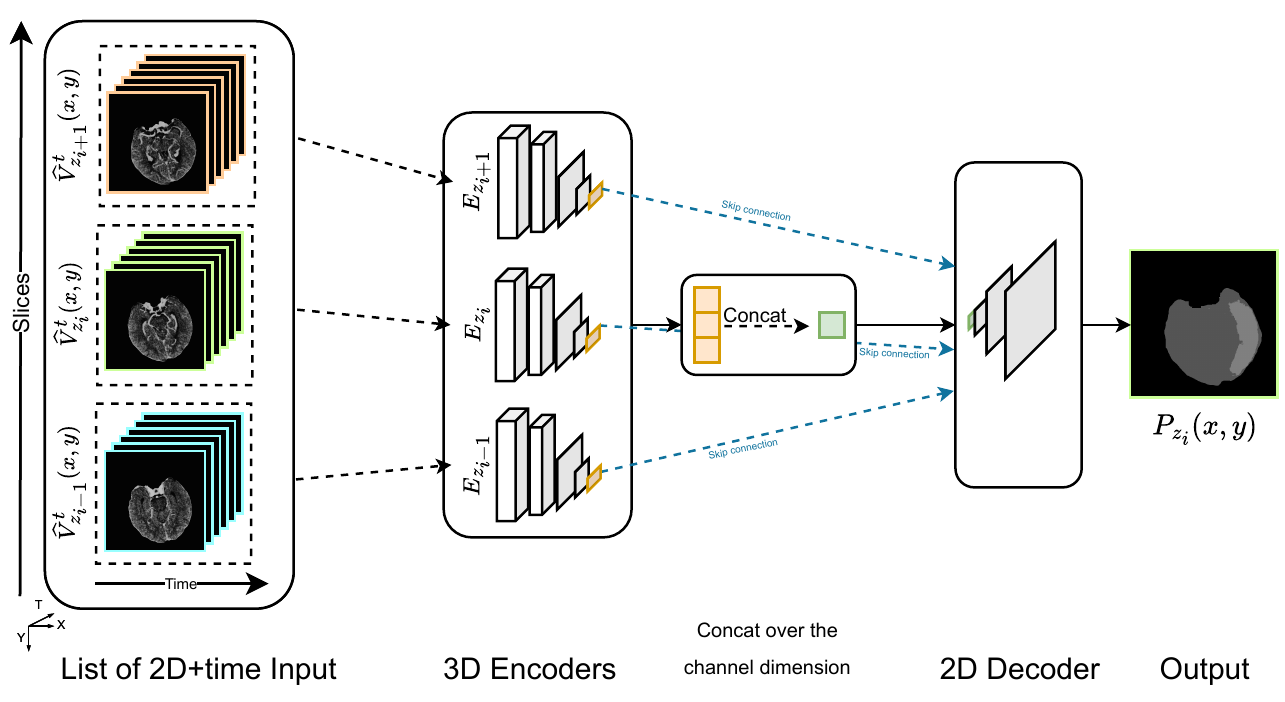}}
\subcaption{\emph{3D+time mJ-Net} architecture (Sec. \ref{3dmj})}
\label{fig:comparison:3d}
\end{minipage}
\begin{minipage}[b]{.9\columnwidth}
\centering
\centerline{\includegraphics[width=\columnwidth]{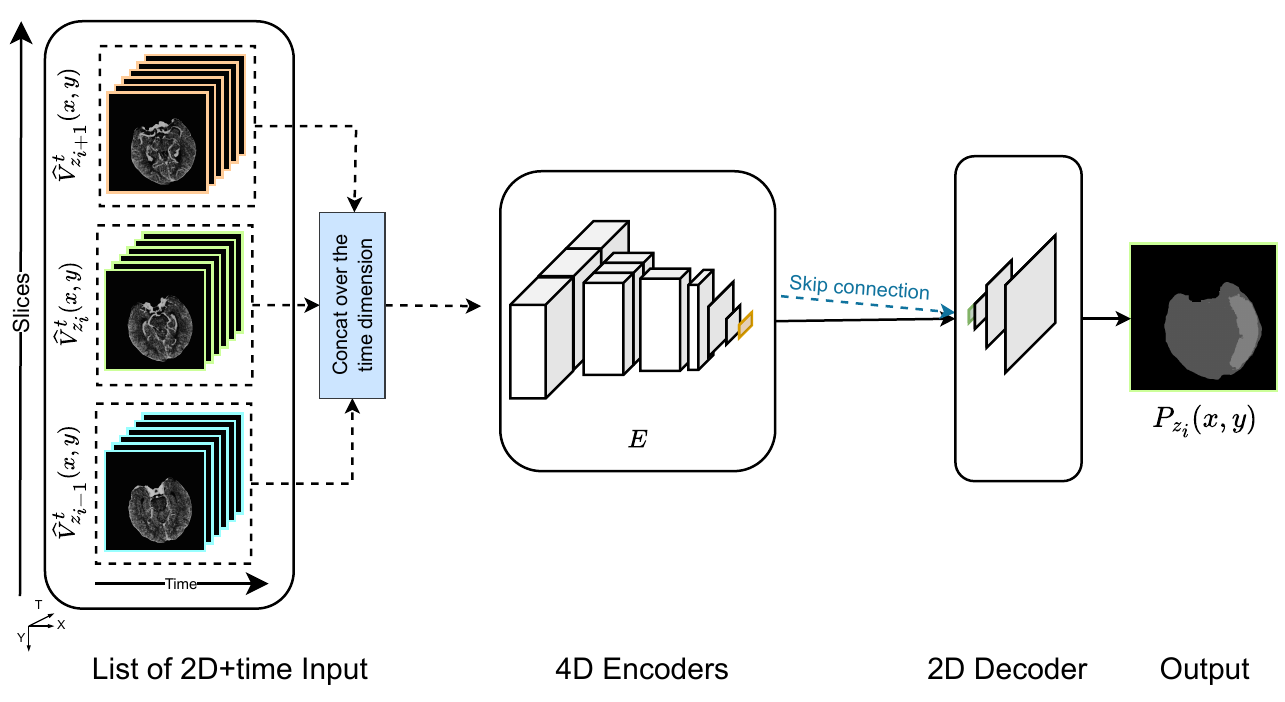}}
\subcaption{\emph{4D mJ-Net} architecture (Sec. \ref{4dmj})}
\label{fig:comparison:4d}
\end{minipage}
\caption{Visual comparison between the proposed architectures: (a) the \emph{3D-TCN}, (b) the \emph{3D+time mJ-Net},  and (c) \emph{4D mJ-Net}.}
\label{fig:comparison}
\end{figure}

\begin{figure*}[h!]
\centering
\centerline{\includegraphics[width=.9\linewidth]{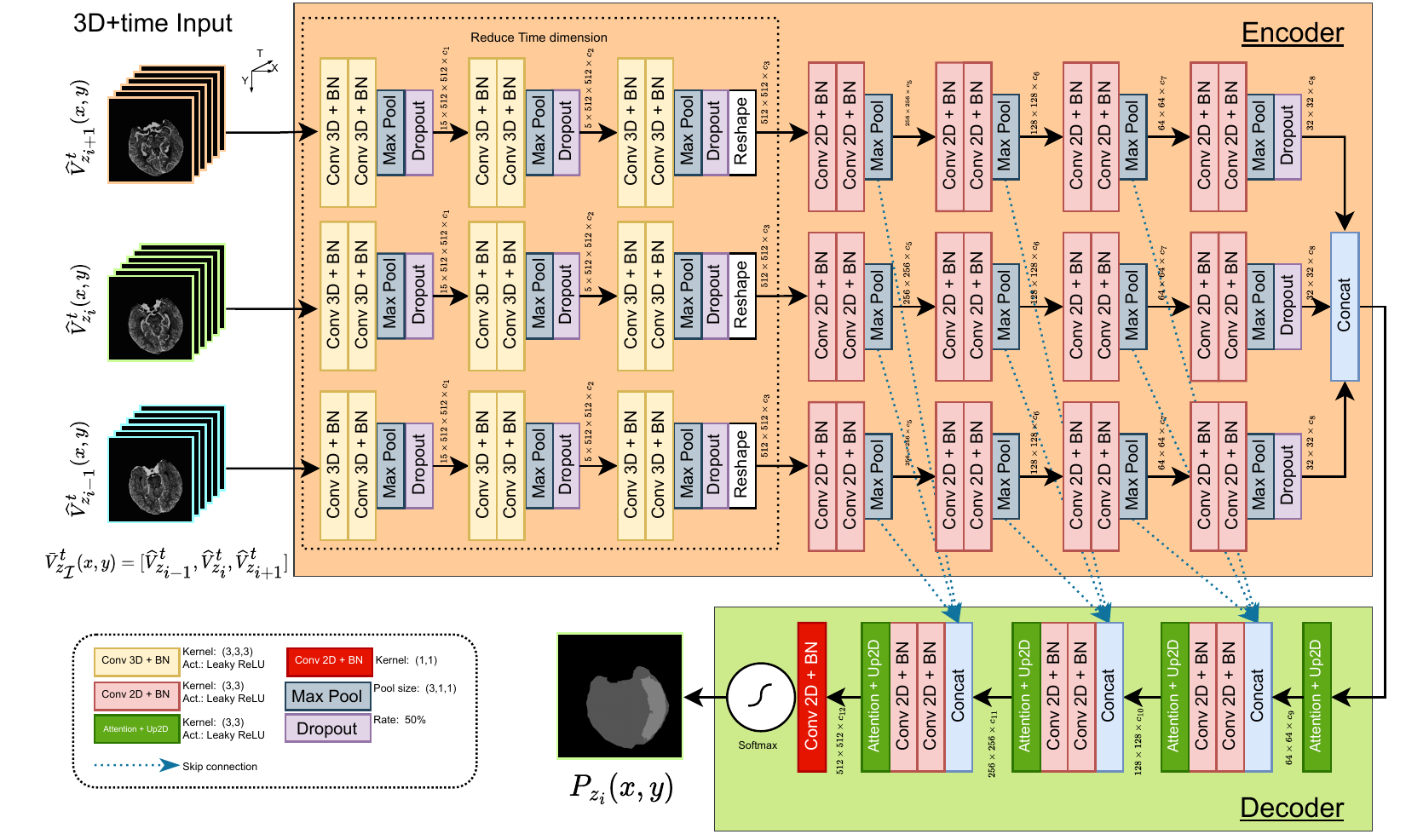}}
\caption{Illustration of the \emph{3D+time mJ-Net} model. The list of 2D+time input $\bar{V}^t_{z_\mathcal{I}}(x,y) = [\widehat{V}_{z_{i-1}}, \widehat{V}_{z_i}, \widehat{V}_{z_{i+1}}]$ is trained in parallel, where $z_\mathcal{I} = \{z_{i-1},z_i,z_{i+1}\}$.
The output is a 2D image $P_{z_i}(x,y)$.
The first max-pooling layer of each block in the convolution section has a pool size of (2,1,1) to reduce the first dimension by a factor of 2.
The second max-pooling layer uses a pool size of (3,1,1), while the third has a pool size of (5,1,1).
The selection of these pool sizes is due to reducing the time dimension.
The remaining max-pooling layers have a pool size of (2,1,1).
The Attention layers utilize a kernel of dimension 3 and a Leaky ReLU activation function.
The 2D Upsampling layers have an upsampling factor of 2.
The last convolution layer has a kernel of 1 and a Softmax activation function to produce a probability score for every class.
}
\label{fig:3darch}
\end{figure*}

\subsection{Proposed 4D methods}\label{sec:propmet}

All the proposed methods\footnote{The code is publicly available at \url{https://github.com/Biomedical-Data-Analysis-Laboratory/4D-mJ-Net}} use the entire 4D CTP scan as input; the main difference lies in how the 4D input is processed.
The \emph{3D-TCN} is based on a \emph{2D-TCN} \citep{amador2021stroke}, modified to receive a list of 3D input volumes.
The \emph{3D+time mJ-Net} inputs a list of 2D+time brain volumes from a CTP dataset, while the \emph{4D mJ-Net} uses the entire 4D structure of a CTP dataset as input.
Fig. \ref{fig:comparison} compares these architectures with their respective inputs.

\subsubsection{Approach 5: 3D Temporal Convolutional Network}\label{3dtcn}

We extend the architecture proposed by \citet{amador2021stroke} for our application to exploit further the information in the depth dimension.
In the remainder of the paper, we call our architecture \emph{3D-TCN}.
The main differences between the proposed \emph{3D-TCN} and the \emph{3D-TCN-SE} (Sec. \ref{sec:3dtcnse}) rely on the usage of a 3D encoder for each input element, instead of a 3D single encoder, plus the possibility to segment both core and penumbra regions, in comparison with segmenting only the core areas.

The 4D CTP scans for the \emph{3D-TCN} are all re-sampled to 1 second per time point. 
As described in Sec. \ref{tcn}, the \emph{3D-TCN} architecture feeds each element of the input list $\widehat{V}^{t_j}_{z_\mathcal{I}}$ at time point $t_j$ to a specific 3D encoder $E^{t_j}_{\text{3D-TCN}}$ to extract low-level features.
Each $E^{t_j}_{\text{3D-TCN}}$ encoder returns a $(4\times4\times C)$ feature vector, where $C$ corresponds to the number of channels.
Each feature vector is merged to create a single input $\text{ETOT}_{\text{3D-TCN}} = [E^{t_j}_{\text{3D-TCN}} | \forall t_j \in t]$.
The $\text{ETOT}_{\text{3D-TCN}}$ is used in the TCN, which generates a one-dimensional vector $O_{\text{3D-TCN}}$ of 64 elements.
The TCN's output $O_{\text{3D-TCN}}$ is then given in input to the decoder to create the final predicted 2D image $P_{z_i}(x,y)$ of a brain slice $z_i$ at index $i$.

\begin{figure*}[ht!]
\centering
\centerline{\includegraphics[width=\linewidth]{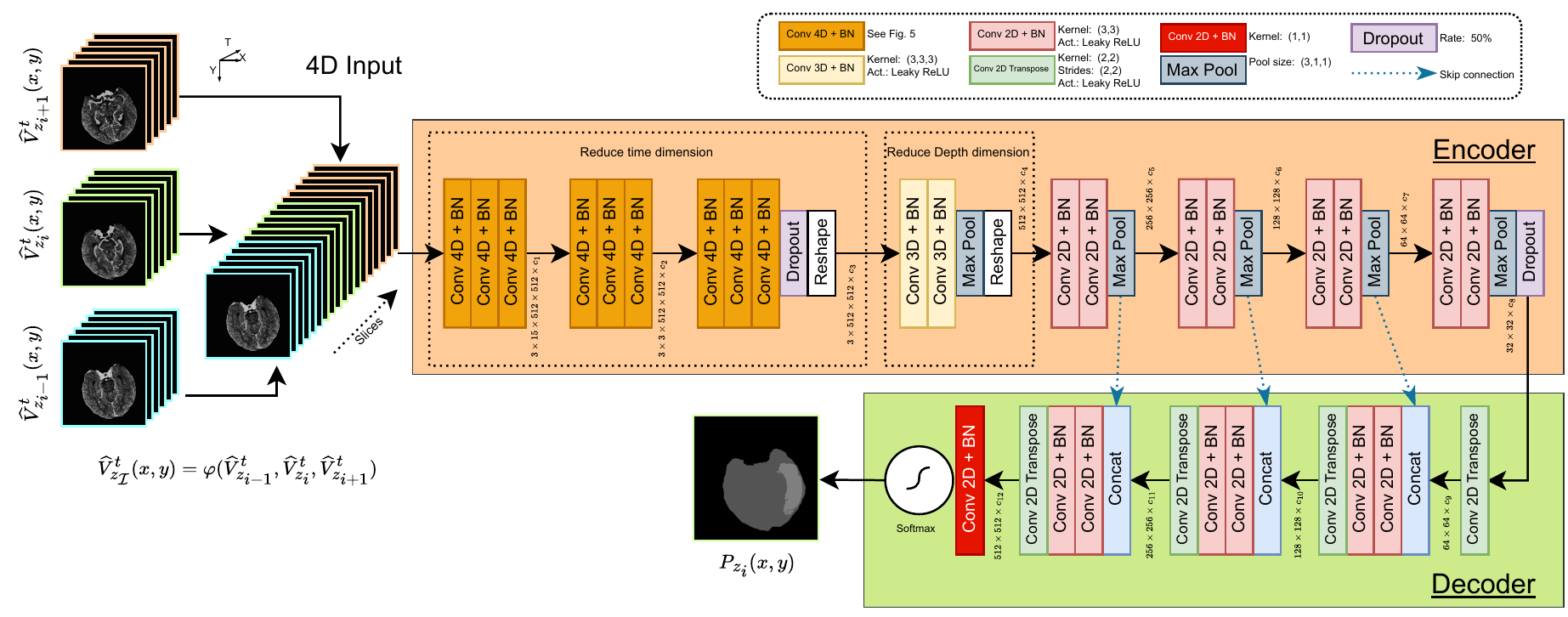}}
\caption{Illustration of the \emph{4D mJ-Net} architecture.
The 4D input $\widehat{V}^t_{z_\mathcal{I}} = \varphi(\widehat{V}_{z_{i-1}}^t,\widehat{V}_{z_i}^t,\widehat{V}_{z_{i+1}}^t)$ is the concatenation of a 2D+time volume $\widetilde{V}^t_{z_{i}}$ of a brain slice $z_i$ at index $i$ over all the time points $t$ plus its neighbouring brain slice volumes ($\widetilde{V}^t_{z_{i-1}},\widetilde{V}^t_{z_{i+1}}$).
Two MonteCarlo dropout layers \citep{gal2016dropout} are added at the end of the 4D and 2D Convolution blocks. The rate was set to $50\%$.
These layers were added to reduce uncertainties in the final predictions.
The last convolution layer has a kernel of ($1\times1$) and a Softmax activation function to produce a probability score for every class.
}
\label{fig:arch}
\end{figure*}

\subsubsection{Approach 6: 3D+time mJ-Net}\label{3dmj}

We propose a model called \emph{3D+time mJ-Net}, an extension of the work of \citet{tomasetti2020cnn}.
The proposed model inputs a list of 2D+time matrices; thus, the dimension of this input can be defined as 3D+time.
The input and output are presented in Table \ref{tab:methods}, whereas a visual example of the input for the model is given in Fig. \ref{fig:inputs}.
Each element of the input list coincides with a possible input for the \emph{mJ-Net} (details in Sec. \ref{25dmj}).
$\bar{V}_{z_\mathcal{I}}^t$ consists of a list of 2D+time volumes, where $z_\mathcal{I} = \{z_{i-1},z_i,z_{i+1}\}$ is a set of brain slices containing the $i$th slice $z_i$ analyzed and its neighboring slices $z_{i-1}$ and $z_{i+1}$.
In case the index $i$ corresponds to the first (or last) brain slice, $\widehat{V}^t_{z_{i-1}}$ (and equivalently $\widehat{V}^t_{z_{i+1}}$) is set equal to $\widehat{V}^t_{z_{i}}$.
Every 2D+time volume from the input list $\widehat{V}_{z_i}^t(x,y)$ is trained separately in the model through a series of encoders ($E_{z_{i-1}}, E_{z_{i}}, E_{z_{i+1}}$) composed of 3D-Conv and 2D-Conv layers.
Each section is independent of the other: the convolution layers have no shared weights.

Fig. \ref{fig:3darch} illustrates the model's architecture.
Attention layers \citep{oktay2018attention} and 2D Upsampling layers were implemented in the decoder section.
Attention layers benefit the architecture by focusing on target structures and help increase the segmentation performances. 
With the sole exception of the last convolution layer, each convolution layer uses a kernel of dimension 3 and a Leaky ReLU activation function \citep{graham2014spatially} with $\alpha=1/3$.

\subsubsection{Approach 7: 4D mJ-Net}\label{4dmj}

We propose another model called \emph{4D mJ-Net}.
We introduce this method to avoid the three paths to process the 4D data presented in the previous architecture (Sec. \ref{3dmj}).
We still use a sliding window technique over the depth dimension to simultaneously limit the amount of input data fed to the model, using three consecutive brain slices at a time.
Like the \emph{3D+time mJ-Net} model, also this approach is an extension of the work of \citet{tomasetti2020cnn}.
Information on this approach is given in Table \ref{tab:methods}.
The 4D input tensor $\widehat{V}^t_{z_\mathcal{I}}$ contains both the time dimension and the neighboring slices of the $i$th brain slice.
The $\widehat{V}^t_{z_\mathcal{I}}$ is a concatenation of a 2D+time volume $\widehat{V}^t_{z_{i}}$ of a brain slice $z_i$ at index $i$ over all the time points $t$ together with its neighbouring 2D+time volumes $\widehat{V}^t_{z_{i-1}},\widehat{V}^t_{z_{i+1}}$.
This model can be considered an early-fusion approach since the 4D input tensors $\widehat{V}_{z_{i-1}}^t,\widehat{V}_{z_i}^t,\widehat{V}_{z_{i+1}}^t$ are concatenated before being fed to the encoder's model.

The proposed \emph{4D mJ-Net} model is a combination of both \emph{3D+time mJ-Net} (Sec. \ref{3dmj}) and \emph{mJ-Net} (Sec. \ref{25dmj}).
The proposed approach uses the same input type that the \emph{3D+time mJ-Net} exploits.
However, rather than a list of 2D+time volumes, the model concatenates the input into a single 4D tensor of dimensions ($X \times Y \times Z \times T$).

Unlike 1D, 2D, and 3D Convolution layers, 4D Convolution layers are not available in public DL frameworks (i.e., Keras\footnote{\url{https://keras.io/}} or PyTorch\footnote{\url{https://pytorch.org/}}).
Thus, for this model, we implemented a novel 4D-Conv layer (details in Sec. \ref{4dconv}) which uses the convolutional layers defined in the public DL frameworks to replicate a 4D convolution operation.

The architecture of the \emph{4D mJ-Net} is displayed in Fig. \ref{fig:arch}.
No Attention layers \citep{oktay2018attention} were included in these models due to a considerable performance decline.
The output of the 4D-Conv layers is a tensor where the temporal dimension has been squeezed and reduced; information are extrapolated from the temporal dimension.
Thus the output resulting from the 4D-Conv layers contains only three dimensions ($X \times Y \times Z$) plus the channel dimension.
3D-Conv layers are implemented to reduce the depth dimension $Z$ and produce a 2D vector ($X \times Y$) plus the channel dimension.

A weighted categorical cross-entropy (WCC) loss \citep{van2019multiclass} was the loss function implemented for this method.
The loss can be written as:
\begin{align*}
\text{WCC}(x,y) = \sum_c^{\mathcal{C}} \sum_i^{M\times N} (y_{i,c} \log x_{i,c}) \cdot (w_{i,c} y_{i,c})
\end{align*}
where $w_{i,c}$ corresponds to the weight of the $i$th pixel for a class $c \in \mathcal{C}$; $x_{i,c}$ is the $i$th predicted pixel, and $y_{i,c}$ is the corresponding ground truth pixel.


\subsection{Implementation details}\label{details}

\begin{table}[ht]
\caption{Summary of the approaches.}
\label{tab:methods}
\centering
\resizebox{\linewidth}{!}{
\begin{tabular}{c|c|c|c}
\Xhline{3\arrayrulewidth}
\multirow{1}{*}{\textbf{Approach}} & \multirow{1}{*}{\textbf{Input}} & \multirow{1}{*}{\textbf{Output}} & \multirow{1}{*}{\textbf{Loss}}  \\
\Xhline{3\arrayrulewidth}
\makecell{Approach 1 \citep{tomasetti2022multi}: \\ \emph{Multi-input PMs}} & \makecell{$\text{PMs}_{z_i} = [\text{CBF}_{z_i}, \text{CBV}_{z_i}$, \\ $\text{TMAX}_{z_i}, \text{TTP}_{z_i}, \text{MIP}_{z_i}]$} & $P_{z_i}(x,y)$ & FTL \citep{abraham2019novel} \\ \hline 
\makecell{Approach 2 \citep{tomasetti2020cnn}: \\ \emph{mJ-Net}} &  $\widehat{V}_{z_i}^t (x,y) =\varphi(\tilde{I}_{z_i}^{t_j}(x,y) | \forall t_j \in t)$. & \ditto & SDCL \citep{milletari2016v} \\ \hline 
\makecell{Approach 3 \citep{amador2021stroke}: \\ \emph{2D-TCN}} & $\bar{V}^{t}_{z_i} = [ \tilde{I}_{z_i}^{t_j} | \forall t_j \in t ]$ of 2D images $\tilde{I}_{z_i}^{t_j}$ & \ditto & DCL \\ \hline 
\makecell{Approach 4 \citep{amador2022predicting}: \\ \emph{3D-TCN-SE}} & $\bar{V}^{t}_{z_\mathcal{I}} = [\widehat{V}^{t_j}_{z_\mathcal{I}} | \forall t_j \in t] $ & \ditto & SDCL \citep{milletari2016v} \\ \hline 
\makecell{Approach 5: \\ \emph{3D-TCN}} & \ditto & \ditto & \ditto \\ \hline 
\makecell{Approach 6: \\ \emph{3D+time mJ-Net}} & $\bar{V}_{z_\mathcal{I}}^t (x,y) = [\widehat{V}_{z_{i-1}}^t,\widehat{V}_{z_i}^t,\widehat{V}_{z_{i+1}}^t]$ & \ditto & \ditto \\ \hline 
\makecell{Approach 7: \\ \emph{4D mJ-Net}} & $\widehat{V}^t_{z_\mathcal{I}} = \varphi(\widehat{V}_{z_{i-1}}^t,\widehat{V}_{z_i}^t,\widehat{V}_{z_{i+1}}^t)$ & \ditto & WCC \citep{van2019multiclass} \\ \hline 
\end{tabular}
}
\end{table}

\begin{table*}[!ht]
\caption{ \textbf{Experiment results for the validation set}. Values in bold exhibit the best results for each column and each class.
Mean results plus standard deviation for Dice Coefficient (DC), Hausdorff Distance (HD), and $\Delta V$ are presented.
Results are for the penumbra and core areas divided by the distinct patient groups (LVO, Non-LVO, WIS, and All).
Note that for the DC, higher values are better ($\Uparrow$), while for HD and $\Delta V$, lower values are preferable ($\Downarrow$).
}
\centering
\resizebox{\linewidth}{!}{%
\begin{tabular}{c|cccccccccc}
\Xhline{3\arrayrulewidth}
\multirow{2}{*}{\textbf{Method}} & \multicolumn{3}{c|}{\textbf{DC} $\Uparrow$} & \multicolumn{3}{c|}{\textbf{HD (mm)} $\Downarrow$} & \multicolumn{4}{c}{$\Delta V$ \textbf{ (ml)} $\Downarrow$} \\ \cline{2-11}
& \multicolumn{1}{c|}{\textbf{LVO}} & \multicolumn{1}{c|}{\textbf{Non-LVO}} & \multicolumn{1}{c|}{\textbf{All}} & \multicolumn{1}{c|}{\textbf{LVO}} & \multicolumn{1}{c|}{\textbf{Non-LVO}} & \multicolumn{1}{c|}{\textbf{All}} & \multicolumn{1}{c|}{\textbf{LVO}} & \multicolumn{1}{c|}{\textbf{Non-LVO}} & \multicolumn{1}{c|}{\textbf{WIS}} & \multicolumn{1}{c}{\textbf{All}} \\ \cline{2-11}
\Xhline{3\arrayrulewidth}
& \multicolumn{10}{c}{\textbf{Penumbra}} \\ \hline
\emph{Multi-input PMs} \citep{tomasetti2022multi} & \textbf{0.70$\pm$0.1} & 0.27$\pm$0.3 & 0.47$\pm$0.3 &  \multicolumn{1}{|c}{2.9$\pm$0.4} & 1.4$\pm$0.7 & 2.0$\pm$0.8 & \multicolumn{1}{|c}{27.0$\pm$28.6} & 10.0$\pm$15.5 & 9.8$\pm$8.1 & \textbf{19.0$\pm$24.2} \\ \hline
\emph{mJ-Net} \citep{tomasetti2020cnn} & 0.66$\pm$0.2 & 0.39$\pm$0.3 & 0.50$\pm$0.3 & \multicolumn{1}{|c}{2.9$\pm$0.5} & 2.6$\pm$0.6 & 2.7$\pm$0.7 & \multicolumn{1}{|c}{\textbf{25.5$\pm$20.0}} & 24.7$\pm$29.2 & 45.5$\pm$39.1 & 27.2$\pm$26.0 \\ \hline
\emph{2D-TCN} \citep{amador2021stroke} & 0.12$\pm$0.1 & 0.02$\pm$0.0 & 0.07$\pm$0.1  &\multicolumn{1}{|c}{4.1$\pm$0.5} & 3.8$\pm$0.6 & 4.0$\pm$0.6 & \multicolumn{1}{|c}{81.3$\pm$65.6} & 80.6$\pm$57.8 & 131.6$\pm$93.1 & 86.0$\pm$66.5 \\ \hline
\emph{3D-TCN-SE} \citep{amador2022predicting} & 0.25$\pm$0.1 & 0.05$\pm$0.1 & 0.15$\pm$0.1 & \multicolumn{1}{|c}{6.2$\pm$0.5} & 6.7$\pm$0.4 & 6.4$\pm$0.5 & \multicolumn{1}{|c}{497.9$\pm$157.1} & 559.3$\pm$90.4 & 624.6$\pm$118.7 & 533.1$\pm$137.3\\ \hline 
\emph{3D-TCN}  & 0.23$\pm$0.1 & 0.04$\pm$0.1 & 0.14$\pm$0.1 &\multicolumn{1}{|c}{4.3$\pm$0.4} & 4.4$\pm$0.5 & 4.3$\pm$0.5 & \multicolumn{1}{|c}{85.3$\pm$64.0} & 142.7$\pm$51.4 & 164.2$\pm$43.8 & 114.2$\pm$65.3 \\ \hline
\emph{3D+time mJ-Net} &  \textbf{0.70$\pm$0.1} & 0.42$\pm$0.3 & \textbf{0.53$\pm$0.3} & \multicolumn{1}{|c}{2.6$\pm$0.6} & 1.9$\pm$0.8 & 2.2$\pm$0.9  & \multicolumn{1}{|c}{35.1$\pm$36.1} & 18.3$\pm$26.3 & 2.7$\pm$2.6 & 25.7$\pm$32.4 \\ \hline
\emph{4D mJ-Net}  &  0.66$\pm$0.1 & \textbf{0.44$\pm$0.3} & 0.51$\pm$0.3 & \multicolumn{1}{|c}{\textbf{2.3$\pm$0.6}} & \textbf{1.3$\pm$0.7} & \textbf{1.7$\pm$1.0} & \multicolumn{1}{|c}{41.4$\pm$37.2} & \textbf{6.1$\pm$6.3} & \textbf{0.0$\pm$0.0} & 24.3$\pm$32.9\\ \hline
\Xhline{3\arrayrulewidth}
& \multicolumn{10}{c}{\textbf{Core}} \\ \hline
\emph{Multi-input PMs} \citep{tomasetti2022multi} & \textbf{0.37$\pm$0.3} & \textbf{0.21$\pm$0.3} & \textbf{0.28$\pm$0.3} & \multicolumn{1}{|c}{1.2$\pm$0.8} & \textbf{0.4$\pm$0.4} & 0.8$\pm$0.8 &\multicolumn{1}{|c}{9.4$\pm$20.3} & \textbf{0.8$\pm$1.3} & 0.5$\pm$0.5 & 5.3$\pm$15.3  \\ \hline
\emph{mJ-Net} \citep{tomasetti2020cnn} & 0.27$\pm$0.2 & \textbf{0.21$\pm$0.2} & 0.22$\pm$0.2  & \multicolumn{1}{|c}{1.5$\pm$0.7} & 0.8$\pm$0.6 & 1.2$\pm$0.6 & \multicolumn{1}{|c}{\textbf{5.5$\pm$4.9}} & 1.0$\pm$1.2 & 1.0$\pm$1.1 & \textbf{3.4$\pm$4.3} \\ \hline
\emph{2D-TCN} \citep{amador2021stroke} & 0.02$\pm$0.0 & 0.01$\pm$0.0 & 0.01$\pm$0.0 &\multicolumn{1}{|c}{1.9$\pm$0.7} & 1.5$\pm$0.6 & 1.7$\pm$0.7 & \multicolumn{1}{|c}{11.8$\pm$13.3} & 8.1$\pm$8.2 & 11.0$\pm$11.2 & 10.3$\pm$11.4 \\ \hline 
\emph{3D-TCN-SE} \citep{amador2022predicting} & 0.00$\pm$0.0 & 0.00$\pm$0.0 & 0.00$\pm$0.0 &\multicolumn{1}{|c}{1.2$\pm$0.9} & 0.4$\pm$0.4 & 0.8$\pm$0.8 & \multicolumn{1}{|c}{12.7$\pm$15.6} & 1.9$\pm$2.8 & \textbf{0.0$\pm$0.0} & 7.5$\pm$12.8  \\ \hline 
\emph{3D-TCN}  & 0.02$\pm$0.0 & 0.01$\pm$0.0 & 0.01$\pm$0.0  &\multicolumn{1}{|c}{1.4$\pm$0.8} & 0.8$\pm$0.4 & 1.1$\pm$0.7 & \multicolumn{1}{|c}{12.0$\pm$14.3} & 1.9$\pm$2.1 & 2.4$\pm$1.9 & 7.3$\pm$11.6  \\ \hline 
\emph{3D+time mJ-Net} &  0.21$\pm$0.2 & 0.12$\pm$0.2 & 0.16$\pm$0.4 & \multicolumn{1}{|c}{\textbf{1.1$\pm$0.7}} & \textbf{0.4$\pm$0.4} & \textbf{0.7$\pm$0.7} & \multicolumn{1}{|c}{8.1$\pm$10.6} & 1.3$\pm$1.6 & \textbf{0.0$\pm$0.0} & 4.8$\pm$8.5 \\ \hline
\emph{4D mJ-Net} &  0.29$\pm$0.2 & \textbf{0.21$\pm$0.2} & 0.23$\pm$0.2 & \multicolumn{1}{|c}{1.6$\pm$0.9} & 0.5$\pm$0.4 & 1.0$\pm$0.9 & \multicolumn{1}{|c}{25.9$\pm$37.0} & 1.4$\pm$2.2 & \textbf{0.0$\pm$0.0}& 14.3$\pm$29.6 \\ \hline
\end{tabular}%
}
\label{tab:fff}
\end{table*}

Table \ref{tab:methods} provides information about all the methods.
All the methods mentioned in Sec. \ref{sec:exmet} and Sec. \ref{sec:propmet} utilize Adam as the optimizer \citep{kingma2014adam} with a learning rate of 0.0003 and a step-based decay rate of 0.95 every ten epochs.
The batch size is set to 2. 
An early stopping function is called if there is no decrement in the validation loss after 25 epochs.
During training, L1 and L2 regularizations are applied in the kernels, plus a max norm constraint is also used in the kernel and bias weights.
All experiments were implemented in Python using Keras (2.3.1) with Tensorflow as the backend and trained using an NVIDIA Tesla V100 GPU (32 GB memory).

\section{Experiments \& Results}\label{res}
We assess the proposed methods on a local dataset of CTP scans from 152 patients (Sec. \ref{material}).
All experiments are performed with the same training set and evaluated over the validation set (details in Table \ref{tab:division}).
The test set is used only to make predictions with the best models with CTP scans that the methods have not seen before.
Since the Non-LVO group has smaller ischemic areas than the LVO patients, we set a higher penalty for every misclassification of penumbra and core classes for this sub-group during training.

\subsection{Evaluation metrics}
Three evaluation metrics are used to assess the various experiments' models.
The Dice Coefficient ($\text{DC}$), the Hausdorff Distance (HD) \citep{birsan2005one}, and the absolute difference in the volumes ($\Delta V$).
We employ the DC to compare the model predictions with the ground truth segmentations.
The DC  between two segmentations $x$ and $y$ is given by the following equation:
\begin{align*}
\text{DC}(x,y) = 2\frac{|x \cap y|}{|x|+|y|} 
\end{align*}
where the range for the DC is $[0, 1]$; thus a $\text{DC}(x,y)=1$ corresponds to a perfect match between the prediction $x$ and ground truth $y$ segmentations.

The HD measures how two subsets $(\mathcal{A}, \mathcal{B})$ are distant from each other, and it is formulated as follows:
\begin{align*}
\text{HD}(\mathcal{A},\mathcal{B}) = \max\left\{\, h(\mathcal{A},\mathcal{B}),\, h(\mathcal{B},\mathcal{A}) \,\right\}
\end{align*}
where $h(\mathcal{A},\mathcal{B}) = \max_{a \in \mathcal{A}} \min_{b \in \mathcal{B}} ||a - b||$. The range value for the HD is $[0, \infty]$.

The absolute difference in the volumes $\Delta V$ between the prediction volume $V_x$ and the ground truth volume $V_y$ can be expressed as:
\begin{align*}
\Delta V(V_y,V_x) = |V_y- V_x|
\end{align*}
The range for $\Delta V$ is $[0, \infty]$, and $\Delta V(V_y,V_x) = 0$ represents a perfect match between the two volumes.
The $\Delta V(V_y, V_x)$ is an essential evaluation metric for the WIS group due to the lack of ground truth segmentations in this group.
The other metrics are not suitable for understanding how the predictions will be since the ground truth will always be empty.

The best scenario for a model is to produce high DC with low HD and $\Delta V$: this implies a strong correlation between the predicted areas and the ground truth regions.
If the results show high $\Delta V$ (or HD) with low DC, an over-segmentation of the ischemic areas is perceived.
On the other hand, promising outcomes of $ \Delta V $ (or HD) with mediocre DC results imply an under-segmentation of the predicted regions.

\begin{table}[]
\caption{Ablation study for the \emph{4D mJ-Net} model showing how various pre-processing steps (HE, $\gamma$, $z$) and re-sampling ($\uplus$) affect the Dice Coefficient (DC) for the validation results. Penumbra and core DC scores are shown for all the classes together.
Note that for the DC, higher values are better ($\Uparrow$).}
\tiny
\centering
\resizebox{\columnwidth}{!}{%
\begin{tabular}{cccc|cc|cc}
\Xhline{3\arrayrulewidth}
\multicolumn{4}{c|}{\multirow{2}{*}{\textbf{Ablation Setting}}} & \multicolumn{4}{c}{\textbf{DC} $\Uparrow$} \\ \cline{5-8}
& & & & \multicolumn{2}{c|}{\textbf{LVO}} & \multicolumn{2}{c}{\textbf{Non-LVO}} \\ \hline
\textbf{HE} & $\gamma$ & $z$ & $\uplus$ & \textbf{Penumbra} & \textbf{Core} & \textbf{Penumbra} & \textbf{Core} \\
\Xhline{3\arrayrulewidth}
\xmark & \xmark & \xmark & \xmark & 0.42$\pm$0.2 & 0.25$\pm$0.2 & 0.20$\pm$0.2 & 0.16$\pm$0.2 \\ \hline 
\cmark & \xmark & \xmark & \xmark & 0.32$\pm$0.3 & 0.24$\pm$0.2 & 0.13$\pm$0.2 & 0.16$\pm$0.3 \\ \hline
\xmark & \cmark & \xmark & \xmark & 0.00$\pm$0.0 & 0.07$\pm$0.1 & 0.00$\pm$0.0 & 0.06$\pm$0.1  \\ \hline
\cmark & \cmark & \xmark & \xmark & 0.48$\pm$0.2 & 0.28$\pm$0.2 & 0.24$\pm$0.2 & 0.20$\pm$0.3 \\ \hline
\xmark & \xmark & \cmark & \xmark & 0.01$\pm$0.0 & 0.14$\pm$0.2 & 0.01$\pm$0.0 & 0.08$\pm$0.1 \\ \hline
\cmark & \xmark & \cmark & \xmark & 0.53$\pm$0.2 & 0.22$\pm$0.2 & 0.35$\pm$0.3 & 0.12$\pm$0.2 \\ \hline
\xmark & \cmark & \cmark & \xmark & 0.28$\pm$0.2 & 0.17$\pm$0.2 & 0.08$\pm$0.1 & 0.05$\pm$0.1 \\ \hline
\cmark & \cmark & \cmark & \xmark & \textbf{0.66$\pm$0.1} & \textbf{0.29$\pm$0.2} & \textbf{0.44$\pm$0.3} & \textbf{0.21$\pm$0.2} \\ \hline
\xmark & \xmark & \xmark & \cmark & 0.00$\pm$0.0 & 0.00$\pm$0.0 & 0.00$\pm$0.0 & 0.00$\pm$0.0  \\ \hline 
\cmark & \xmark & \xmark & \cmark & 0.07$\pm$0.1 & 0.26$\pm$0.2 & 0.01$\pm$0.0 & 0.05$\pm$0.1 \\ \hline
\xmark & \cmark & \xmark & \cmark & 0.00$\pm$0.0 & 0.00$\pm$0.0 & 0.00$\pm$0.0 & 0.00$\pm$0.0  \\ \hline
\cmark & \cmark & \xmark & \cmark & 0.41$\pm$0.3 & 0.29$\pm$0.2 & 0.11$\pm$0.2 & 0.12$\pm$0.2 \\ \hline
\xmark & \xmark & \cmark & \cmark & 0.00$\pm$0.0 & 0.00$\pm$0.0 & 0.00$\pm$0.0 & 0.00$\pm$0.0  \\ \hline
\cmark & \xmark & \cmark & \cmark & 0.56$\pm$0.2 & 0.24$\pm$0.2 & 0.37$\pm$0.3 & 0.18$\pm$0.2 \\ \hline
\xmark & \cmark & \cmark & \cmark & 0.00$\pm$0.0 & 0.00$\pm$0.0 & 0.00$\pm$0.0 & 0.00$\pm$0.0  \\ \hline
\cmark & \cmark & \cmark & \cmark & 0.59$\pm$0.2 & \textbf{0.29$\pm$0.2} & 0.40$\pm$0.3 & 0.20$\pm$0.2 \\ \hline
\end{tabular}}
\label{tab:ablation}
\end{table}

\begin{table*}[ht!]
\caption{\textbf{Inter-observer variability results for test set}. Values are presented for the two best-proposed architectures (\emph{3D+time mJ-Net}, \emph{4D mJ-Net}) in relation to manual annotations generated separately by two expert neuroradiologists ($\text{NR}_1$, $\text{NR}_2$) over the test set. An investigation of the inter-observer variability between $\text{NR}_1$ and $\text{NR}_2$ is performed (last row or each class).
Note that for the DC, higher values are better ($\Uparrow$), while for HD and $\Delta V$, lower values are preferable ($\Downarrow$).
}
\centering
\resizebox{\linewidth}{!}{%
\begin{tabular}{c|cccccccccc}
\Xhline{3\arrayrulewidth}
\multirow{2}{*}{\textbf{Method}} & \multicolumn{3}{c|}{\textbf{DC} $\Uparrow$} & \multicolumn{3}{c|}{\textbf{HD (mm)} $\Downarrow$} & \multicolumn{4}{c}{$\Delta V$ \textbf{(ml)} $\Downarrow$} \\ \cline{2-11}
& \multicolumn{1}{c|}{\textbf{LVO}} & \multicolumn{1}{c|}{\textbf{Non-LVO}} & \multicolumn{1}{c|}{\textbf{All}} & \multicolumn{1}{c|}{\textbf{LVO}} & \multicolumn{1}{c|}{\textbf{Non-LVO}} & \multicolumn{1}{c|}{\textbf{All}} & \multicolumn{1}{c|}{\textbf{LVO}} & \multicolumn{1}{c|}{\textbf{Non-LVO}} & \multicolumn{1}{c|}{\textbf{WIS}} & \textbf{All} \\ \cline{2-11}
\Xhline{3\arrayrulewidth}
& \multicolumn{10}{c}{\textbf{Penumbra}} \\ \hline
\makecell{\emph{3D+time mJ-Net}\\ vs ($\text{NR}_1$ vs $\text{NR}_2$)} &  0.70$\pm$0.1 & 0.32$\pm$0.3 & 0.51$\pm$0.3 &\multicolumn{1}{|c}{2.6$\pm$0.5} & 1.5$\pm$0.6 & 2.1$\pm$0.9 & \multicolumn{1}{|c}{36.7$\pm$36.4} & 6.5$\pm$4.9 & 10.1$\pm$3.1 & 24.2$\pm$31.1 \\ \hline
\makecell{\emph{4D mJ-Net} \\ vs ($\text{NR}_1$ vs $\text{NR}_2$)} & 0.67$\pm$0.1 & 0.25$\pm$0.3 & 0.47$\pm$0.3 &\multicolumn{1}{|c}{2.4$\pm$0.5} & 0.9$\pm$0.5 & 1.7$\pm$1.0 & \multicolumn{1}{|c}{34.1$\pm$30.6} & \textbf{5.3$\pm$6.6} & \textbf{0.0$\pm$0.0} & 21.4$\pm$27.7 \\ \hline
$\text{NR}_1$ vs $\text{NR}_2$ & \textbf{0.78$\pm$0.1}& \textbf{0.65$\pm$0.2} & \textbf{0.67$\pm$0.2} & \multicolumn{1}{|c}{\textbf{2.2$\pm$0.4}} & \textbf{0.8$\pm$0.6} & \textbf{1.6$\pm$1.0} & \multicolumn{1}{|c}{\textbf{33.3$\pm$27.7}} & 5.5$\pm$9.2 & \textbf{0.0$\pm$0.0} & \textbf{21.0$\pm$25.9} \\ \hline
\Xhline{3\arrayrulewidth}
& \multicolumn{10}{c}{\textbf{Core}} \\ \hline
\makecell{\emph{3D+time mJ-Net}\\ vs ($\text{NR}_1$ vs $\text{NR}_2$)} &  0.19$\pm$0.2 & 0.01$\pm$0.0 & 0.12$\pm$0.2 &\multicolumn{1}{|c}{1.6$\pm$0.8} & 0.3$\pm$0.4 & 1.0$\pm$1.0 & \multicolumn{1}{|c}{14.6$\pm$18.2} & 0.9$\pm$2.3 & \textbf{0.0$\pm$0.0} & 8.7$\pm$15.4 \\ \hline
\makecell{\emph{4D mJ-Net} \\ vs ($\text{NR}_1$ vs $\text{NR}_2$)} & 0.28$\pm$0.2 & 0.03$\pm$0.1 & 0.18$\pm$0.2 &\multicolumn{1}{|c}{1.8$\pm$0.8} & 0.3$\pm$0.5 & 1.1$\pm$1.0 & \multicolumn{1}{|c}{21.2$\pm$31.7} & 1.8$\pm$4.5 & \textbf{0.0$\pm$0.0} & 12.8$\pm$25.9 \\ \hline
$\text{NR}_1$ vs $\text{NR}_2$ & \textbf{0.44$\pm$0.2} & \textbf{0.15$\pm$0.2} & \textbf{0.30$\pm$0.3} & \multicolumn{1}{|c}{\textbf{1.4$\pm$0.6}} & \textbf{0.2$\pm$0.4} & \textbf{0.9$\pm$0.8} & \multicolumn{1}{|c}{\textbf{5.6$\pm$4.3}} &\textbf{ 0.7$\pm$1.9} & \textbf{0.0$\pm$0.0} & \textbf{3.5$\pm$4.2} \\ \hline
\end{tabular}}
\label{tab:iovr}
\end{table*}

\subsection{Comparison with other methods}
The proposed \emph{3D+time mJ-Net}, \emph{4D mJ-Net}, and \emph{3D-TCN} methods are compared with alternative models: the \emph{2D-TCN} \citep{amador2021stroke}, the \emph{mJ-Net} \citep{tomasetti2020cnn}, the \emph{3D-TCN-SE} \citep{amador2022predicting}, and the \emph{Multi-input PMs} \citep{tomasetti2022multi}.

Table \ref{tab:fff} presents the evaluation metrics' results over the validation set.
Results are presented for each group distinctly (LVO, Non-LVO, WIS, and all) to highlight the strengths and weaknesses of each model over the various groups composing the dataset.
An extensive number of experiments are performed for all the analyzed models.
However, to present a fair comparison among the various models, we only introduce the methods with a combination of parameters that yield the best results omitting the other combinations tested during experiments.
Qualitative comparison results of random brain slices extracted from the validation set are provided in Fig. \ref{fig:results}.

\subsection{Ablation Study}\label{sec:ablation}
To demonstrate the effects of the pre-processing steps (Sec. \ref{sec:preproc}), we conduct an ablation study on the \emph{4D mJ-Net} architecture.
Moreover, we re-sampled the CTP scans to handle the irregular temporal dimension and studied the effect of using re-sampled scans during the model's training.
Different CT scan vendors have different imaging acquisition protocols, thus re-sampling the scans to a fixed time-sampling-rate is an reasonable step to increase the versatility and usability across hospitals.
DC are illustrated in Table \ref{tab:ablation} showing performances of the network for all the groups trained with the datasets using different types of pre-processing steps and re-sampled scans.
The study aims to systematically analyze the contribution of each pre-processing step toward improving the overall results.
We begin by defining a baseline configuration consisting of the raw input images without pre-processing (first row in Table \ref{tab:ablation}).
Subsequently, we incrementally introduce and evaluate individual pre-processing steps, such as histogram equalization (HE), gamma correction $(\gamma)$, and z-score ($z$).

\subsection{Inter-observer variability}
Two expert neuroradiologists ($\text{NR}_1$, $\text{NR}_2$) manually annotated the scans of 33 randomly selected patients: 19 from the LVO group, 11 from the Non-LVO, and 3 from the WIS subset.
The manual annotation images were generated using the same criteria endorsed for creating the ground truth images, as explained in Sec. \ref{gt}.
An investigation of the inter-observer variability between $\text{NR}_1$, $\text{NR}_2$, and the two best-proposed models is presented in Table \ref{tab:iovr}.

\begin{figure*}[h!]
\centering
\centerline{\includegraphics[width=\linewidth]{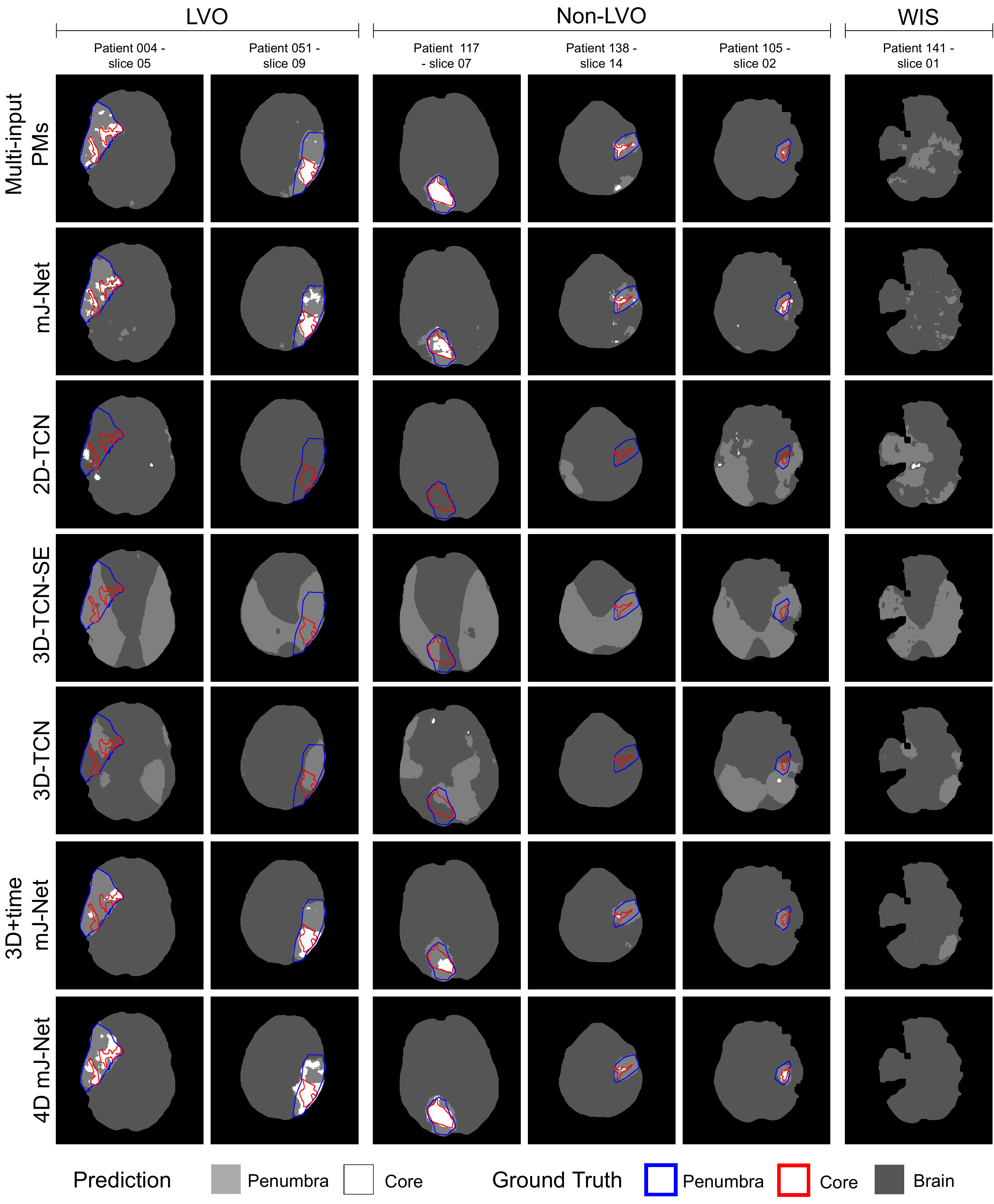}}
\caption{Qualitative comparisons for the tested models. The brain slices are taken randomly from distinct patients from the validation set, divided by group (left to right). Each row represents the results for each model involved in the study.}
\label{fig:results}
\end{figure*}

\section{Discussion}\label{disc}
Early detection and intervention in AIS patients are of vital importance \citep{advani2017golden, meretoja2014stroke, meretoja2017endovascular}. 
In this study, we have proposed different architectures to utilize the 4D CTP input to use the spatio-temporal information better than in existing approaches.
We suggest expanding the \emph{mJ-Net} and showing two ways of segmenting ischemic areas in patients suspected of AIS.
In addition, we expand another method (\emph{3D-TCN}) for comparison reasons.
We use the entire raw 4D CTP data and feed different combinations as input to our proposed approaches to prevent possible loss of spatio-temporal information. Studying the data as an independent volume and neglecting its spatio-temporal nature can lead to the loss of relevant information. All proposed approaches return a series of 2D segmented images as output, later stacked together to produce a 3D volume. Returning a list of 2D images as output is less computationally expensive and less memory intensive than directly returning 3D volumetric data as output \citep{singh20203d}.

Few studies have adopted 4D datasets in DNN models to detect ischemic lesions in patients affected by a stroke \citep{soltanpour2022using, amador2021stroke, robben2020prediction, amador2022predicting}. This is rooted in the high computational complexity of 4D data and the lack of ground truth for the whole set.
The limitations that these approaches encounter are as follows: 1) datasets used for the training and evaluation take into account only a subset of the entire population; 2) segmentations are only performed on the core areas, excluding penumbra regions; 3) ground truth images derived from follow-up DWI or NCCT present some limitations \citep{schellinger2010evidence,goyal2020challenging}.

To our knowledge, this is the first study using 4D CTP data to segment both the ischemic regions, penumbra and core. 
Additionally, we include data from all patients, regardless of stroke severity, to train our models. 
Rather than entrusting ground truth images from follow-up DWI or NCCT studies that are usually taken 24 hours or several days after the onset of stroke, our proposed methods were trained with ground truth images obtained from the CTP captured at admission, PMs, and follow-up scans (Sec. \ref{gt}).

We use three evaluation metrics to assess the models' performances: DC, HD, and $\Delta V$ compared with our previously developed algorithms and other state-of-the-art algorithms. 
Results in Table \ref{tab:fff} demonstrate that increasing the input dimension benefits achieving more precise segmentation, especially for the Non-LVO and WIS groups, regardless of the class. Thus, when a smaller portion of the brain is affected, the whole dataset's usage helps achieve better segmentation results. 
The ablation study (Table \ref{tab:ablation}) shows how including the pre-processing steps and not re-sampling the CTP scans helped improve the overall segmentation performances.
It is worth mentioning that using the pre-processing steps and re-sampled CTP scans yields the second best overall results ( last row in Table \ref{tab:ablation}), establishing the validity of the pre-processing sequence. 

Visual results of random validation brain slices are shown in Fig. \ref{fig:results}, where we can see that our proposed approaches (\emph{3D+time mJ-Net}, \emph{4D mJ-Net}) are less prone to over-segment, especially in the Non-LVO and WIS groups.
It is reported that LVO cases are less common compared to Non-LVO. On average, LVOs are estimated to represent around 30\% of all AIS cases \citep{lakomkin2019prevalence}. 
Thus, a neural network that can accurately segment patients in the Non-LVO group can be valuable in a real-life scenario. 
Nonetheless, patients with LVO represent a clinically significant proportion of patients presenting with AIS, especially considering the grim natural course of the disease.

The results presented in Table \ref{tab:fff} indicate that all \emph{mJ-Net} models have improved where the input data dimension has increased, regardless of the patients' group. 
Fig. \ref{fig:results} shows that using 2D+time input for the \emph{mJ-Net} \citep{tomasetti2020cnn} led to over-segmentation of penumbra class in separate brain tissue sections, brain slice 4-6.
The visual results for the Non-LVO and WIS groups highlight the limitations of this model: the over-segmentation of the penumbra regions might affect the usage in a real-life scenario, and an overestimation of the penumbra area can generate uncertainties for treatment decisions.

Adding depth as an extra dimension to the input of models (\emph{3D+time mJ-Net} and \emph{4D mJ-Net}) determines an increment in the performances for both classes in the three patient groups.
A significant increase is noticeable for the DC metric in the Non-LVO group, regardless of the class.
An essential difference between these two architectures is how they exploit their structures' input.
The \emph{3D+time mJ-Net} is considered a late-fusion approach as the data sources are used independently and fused close to decision-making.
Statistical results presented for the \emph{3D+time mJ-Net} show promising general performances for the LVO group. However, an underestimation of the core class, regardless of the patient group, can be noticed from the visual results in Fig. \ref{fig:results} and the low HD metric.
Nevertheless, \emph{3D+time mJ-Net} achieved the best HD for the core class in all the groups.
The \emph{3D+time mJ-Net} can precisely segment ischemic regions with large areas, as shown by the first three slices in Fig. \ref{fig:results}.
This can also be manifested in the high DC score achieved for the LVO group for both classes (Table \ref{tab:fff}).

The \emph{4D mJ-Net} network has learned to precisely segment the ischemic regions even without re-sampling the CTP scans, as shown in Table \ref{tab:ablation}.
The model fuses the data before they are fed to the network. Visual results in Fig. \ref{fig:results} and values in Table \ref{tab:fff} indicate that the \emph{4D mJ-Net} model segments the penumbra class more precisely compared to the other approaches that use raw CTP as their input. 
This promising performance follows in all patient groups. The \emph{4D mJ-Net} achieved the highest DC metric for core and penumbra regions in patients with Non-LVO. 
This approach gives the best HD for penumbra in all the groups.
The \emph{4D mJ-Net} showed high precision in detecting small ischemic areas, as shown in sample brain slices 3 to 5 in Fig. \ref{fig:results}.
Furthermore, the \emph{4D mJ-Net} model can correctly predict no ischemic regions in WIS patients, as demonstrated by the results for the $\Delta V$ in the WIS group. However, it over-segments the core class in patients with LVO. This means that including the complete spatio-temporal information of the data and following an early fusion approach leads to better prediction in Non-LVO and WIS groups, where small areas are of interest. 

Models based on TCN, in general, showed poor results statistically in Table \ref{tab:fff} and visually in Fig. \ref{fig:results}. They extremely over-segment the penumbra class and poorly segment the core class. 
The original \emph{2D-TCN} and \emph{3D-TCN-SE} were designed to segment only one class, the ischemic core. This can explain the poor performance of segmenting the two classes. 
Besides, in \citet{amador2022predicting} (\emph{3D-TCN-SE}), the model was trained to use only the ipsilateral hemisphere. 
For a fair comparison, the model's training was done over both hemispheres, which can cause over-segmentation in penumbra regions.

As the name indicates, the \emph{Multi-input PMs} model \citep{tomasetti2022multi} takes parametric maps and pre-processed data obtained from CTP scans. The experiment results of this model show a high DC value for the penumbra class in the LVO group, as also seen in the first three brain slices of Fig. \ref{fig:results}.
This highlights that this method presents satisfactory results for large ischemic areas.
However, when the region's volume is small or vacant, the predictions are not optimal: see brain slices 4 and 6 in Fig. \ref{fig:results}.
Although HD and $\Delta V$ are encouraging, DC values show under-segmentation in the core and penumbra classes for the Non-LVO set. 
Using PMs derived from CTP scans limits the machine to only learn from specific pre-processed information.

The inter-observer variability results, highlighted in Table \ref{tab:iovr}, show promising outcomes for the proposed methods with the results achieved by the two expert neuroradiologists ($\text{NR}_1$, $\text{NR}_2$).
Similar statistic values can be observed between $\text{NR}_1$ vs. $\text{NR}_2$ and the \emph{4D mJ-Net} for the penumbra class in connection with the LVO group.
The same results as the neuroradiologists were achieved by the \emph{4D mJ-Net} for the $\Delta V$ in the WIS group.
The proposed \emph{3D+time mJ-Net} model produces higher results for the DC compared to the \emph{4D mJ-Net} in association with the penumbra class. However, the results for the core regions could be more satisfactory.
The inter-observer variability outcomes for the HD and $\Delta V$ highlight substantial similarity among the proposed approaches and the neuroradiologists, except for the core class connected with the LVO group.
The difference can be due to an over-segmentation of this particular class, which can be highly complex to detect for the models considering its small size.

\subsection{Common limitations}
All the assessed approaches have faced general limitations. The images used during the training of each model are from CT scanners of the same vendor. This causes a lack of diversity in the data. 
The annotations used as ground truth surround the essential ischemic regions (penumbra and core) but do not represent the areas perfectly. They might leave out small parts of the core spread into the penumbra tissue and details of the penumbra misclassified as healthy brain tissue \citep{tomasetti2020cnn}.

\section{Conclusions}
Fast and precise diagnosis and treatment are of vital importance in AIS patients.
In this paper, we proposed to use 4D CTP as input to extract spatio-temporal information for segmenting core and penumbra areas in patients with AIS. This is presented primarily by expanding the \emph{mJ-Net} in two ways.
Furthermore, we introduced a novel 4D-Conv layer to exploit spatio-temporal information.
Two of our approaches (\emph{3D+time mJ-Net} and \emph{4D mJ-Net}) achieved promising results for all the classes involved.
\emph{3D+time mJ-Net} can precisely delineates large ischemic areas, while underestimating the core class. 
Our best network (\emph{4D mJ-Net}) can correctly segments penumbra regions, regardless of patient groups with a 0.53 DC score on average.
However, with an average of 0.23 DC score, it overestimates the core class for the LVO group. 

We used the entire 4D CTP dataset of all patients and compared models using different input types. We demonstrated that relying only on images derived from the CTP scans (i.e., PMs) or on a restricted number of dimensions (i.e., 2D, 2D+time, 3D) limits the prediction accuracy in DNN-based approaches.
Moreover, we segmented both penumbra and core regions in ischemic brain tissue since an accurate and fast understanding of both is essential for fast treatment decisions in AIS.

Further studies with larger datasets, including images from different vendors and various acquisition parameters, are still needed to validate our methods. The ISLES18 dataset \citep{hakim2021predicting} can be used in future work; the dataset uses FIAs as ground truth labels, which is not in the scope of the aforementioned architectures, thus some changes must be implemented for validating the methods.
Due to complex and time-consuming work for manual annotations, further work on optimizing the segmentation using unsupervised neural networks is encouraged.


\section*{Acknowledgment}
We declare that we do not have any commercial or associative interest that represents a conflict of interest in connection with the work submitted.

\printbibliography

\begin{IEEEbiography}[{\includegraphics[width=1in,height=1.25in,clip,keepaspectratio]{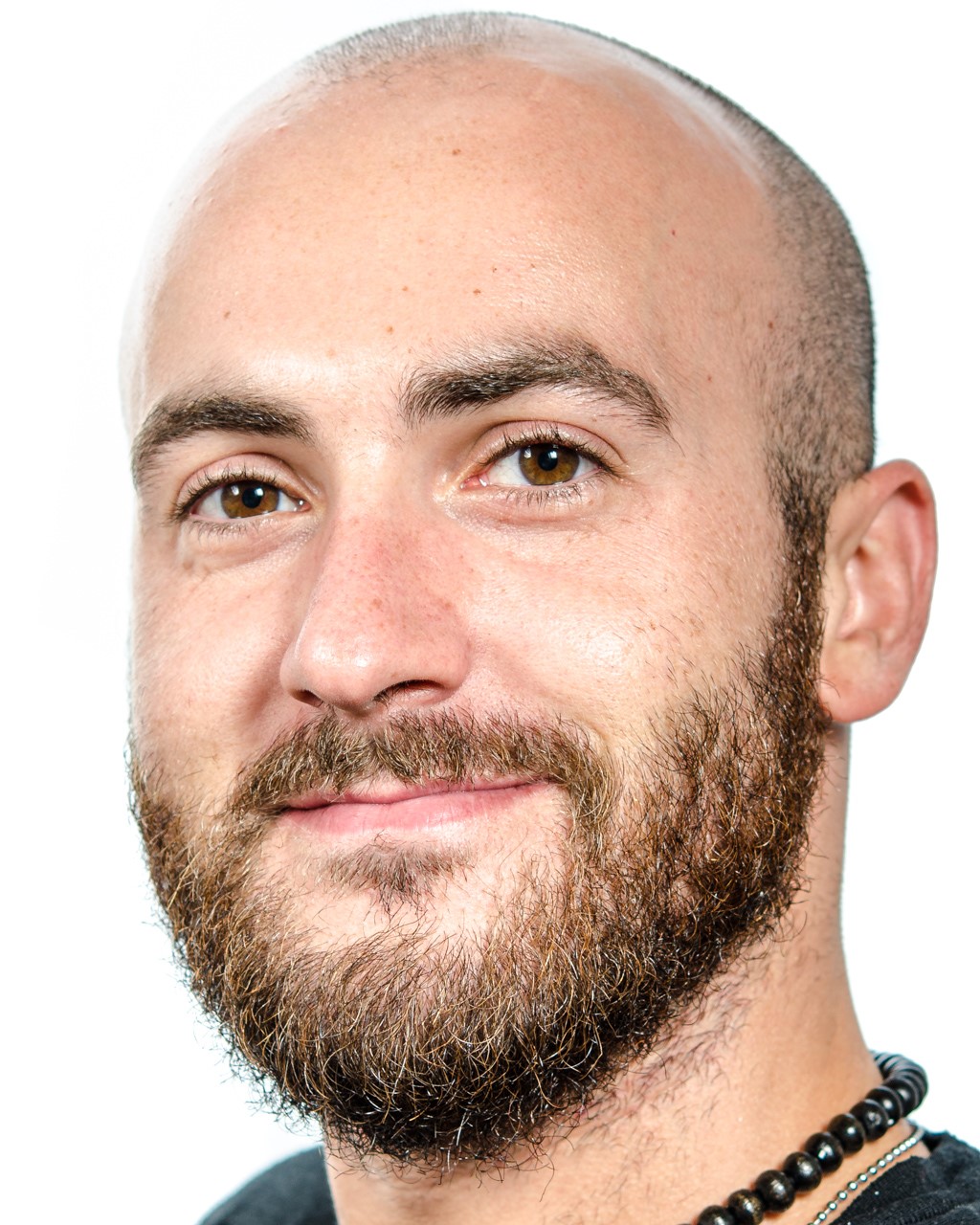}}]{Luca Tomasetti} (M'2021) received the B.S. degree in computer science from Università di Bologna, Bologna, Italy, in 2015 
and the M.S. degree in computer science from University of Stavanger (UiS), Stavanger, Norway, in 2019. He is currently pursuing the Ph.D. degree in artificial intelligence at the Computer Science Department of UiS.
He is part of the project “In depth analysis of perfusion computed tomography (CTP) in patients with acute ischemic stroke” at UiS. The project is part of a twin project performed in close cooperation with a PhD candidate in radiology. 
His research interest are in image processing and machine learning with emphasis on ischemic stroke. 
He is also a member of the Biomedical data analysis laboratory (BMDLab).
\end{IEEEbiography}
\vspace{-2cm}
\begin{IEEEbiography}[{\includegraphics[width=1in,height=1.25in,clip,keepaspectratio]{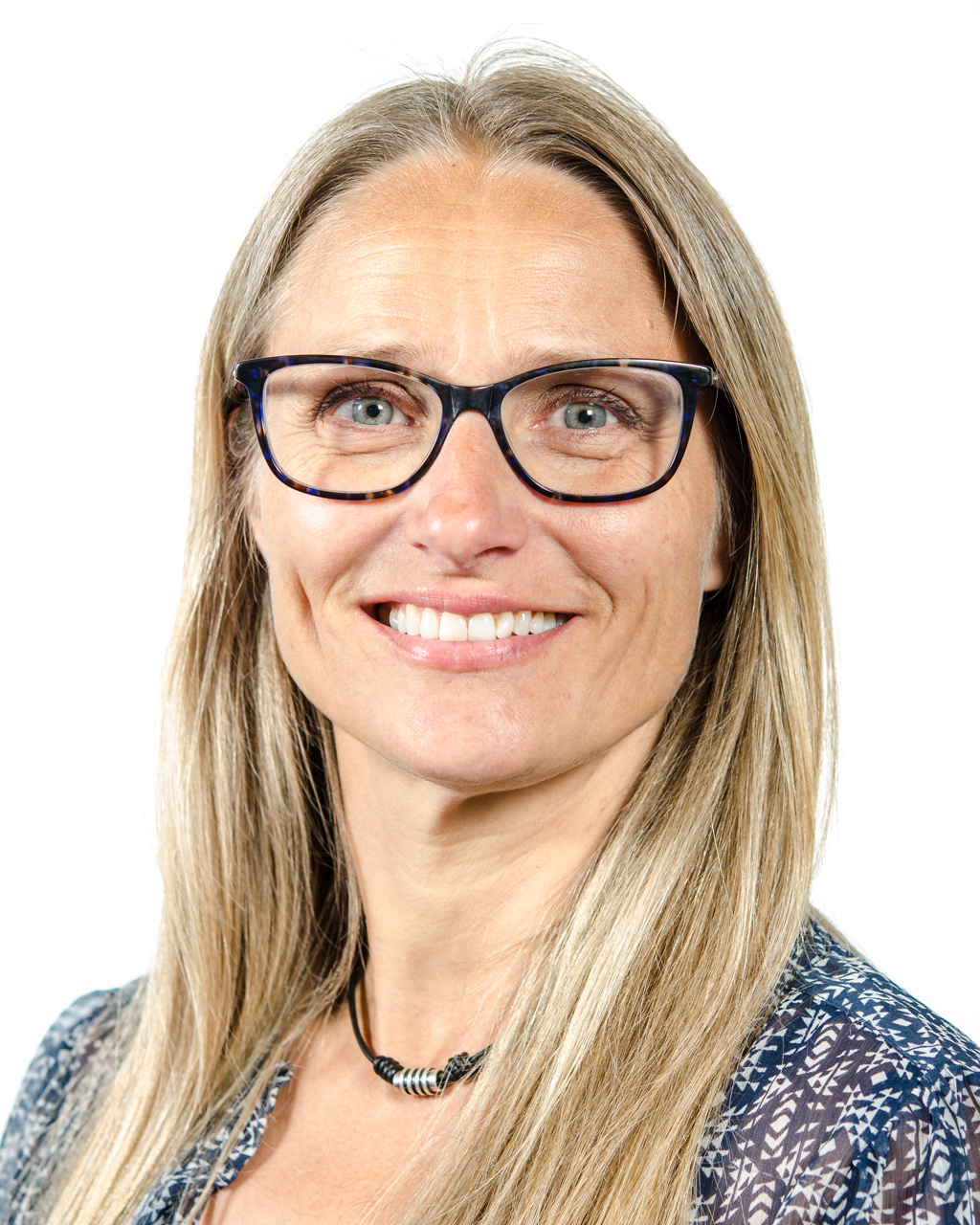}}]{Kjersti Engan} (M'1996, SM'2018) is a professor at the Electrical Engineering and Computer Science department at the University of Stavanger (UiS), Norway. 
She received the BE degree in electrical engineering from Bergen University College in 1994 and the M.Sc. and Ph.D degrees in 1996 and 2000 respectively, in electrical engineering and information technology from the UiS.  
She is the leader of the Biomedical data analysis lab - BMDLab - at UiS.  

 Her research interests are in signal and image processing and machine learning with emphasis on medical applications and in dictionary learning for sparse signal and image representation.  She has a particular interest in AI for newborn survival, stroke detection from CTP imaging and AI in computational pathology.  
 She is a senior member of IEEE.  She has served as Associate editor and Senior Area editor for IEEE Signal Processing Letters and as a member of IEEE Image, Video, and Multidimensional Signal Processing Technical Committee (IVMSP) and associate editor for SIAM Journal on Imaging Sciences (SIIMS).
\end{IEEEbiography}
\vspace{-2cm}
\begin{IEEEbiography}[{\includegraphics[width=1in,height=1.25in,clip,keepaspectratio]{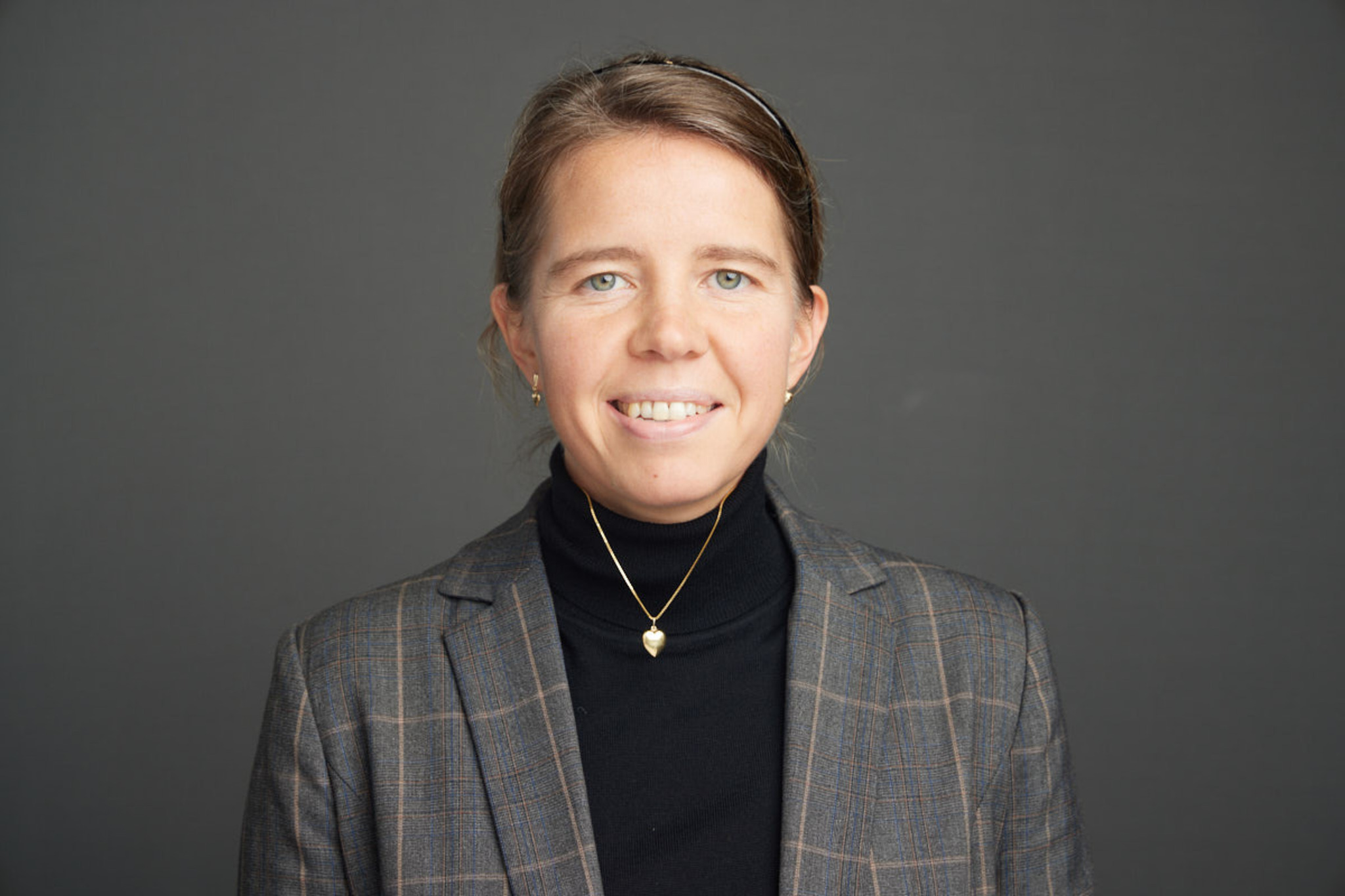}}]{Liv Jorunn H\O{}llesli} (MD, PhD candidate) is at the end of her PhD candidacy on the project ``In depth analysis of perfusion computed tomography (CTP) in patients with acute ischemic stroke'' at the Electrical Engineering and Computer Science Department at the University of Stavanger (UiS), Norway. The project is part of a twin project performed in close cooperation with a PhD student on computer science. 
Dr. H\o{}llesli is also a senior consultant radiologist at Stavanger University Hospital, Norway, holding the European Diploma in Neuroradiology. She received the cand.med.-degree from the Faculty of Medicine, University of Bergen, Norway, in 2012. Neuroradiology, with focus on medical imaging and AI in ischemic stroke, is her primary research focus. H\o{}llesli is a member of the European Society of Neuroradiology, the Norwegian Society of Neuroradiology and the Norwegian Medical association
\end{IEEEbiography}
\vspace{-2cm}
\begin{IEEEbiography}[{\includegraphics[width=1in,height=1.25in,clip,keepaspectratio]{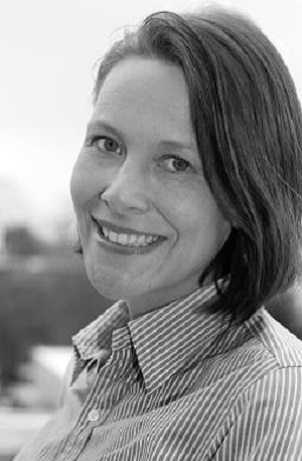}}]{Kathinka D{\AE}hli Kurz}(MD, PhD, Professor II) is a senior consultant radiologist at Stavanger University Hospital, Norway, holding the European Diploma in Neuroradiology. She also has a professor II position at the Electrical Engineering and Computer Science Department at the University of Stavanger (UiS), Norway. She received the cand.med.-degree from the Albert-Ludwigs University of Freiburg, Germany, in 2000. She holds a PhD degree from Germany (2004), and a PhD degree on ``Dynamic MRI: An important tool in problem-solving of breast abnormalities'' from the University of Oslo, Norway (2011). 
Her primary research interests are ischemic stroke, neuroradiology and magnetic resonance imaging. Kurz is a member of the European Society of Neuroradiology, the Norwegian Society of Neuroradiology and the Norwegian Medical association.    
\end{IEEEbiography}
\begin{IEEEbiography}[{\includegraphics[width=1in,height=1.25in,clip,keepaspectratio]{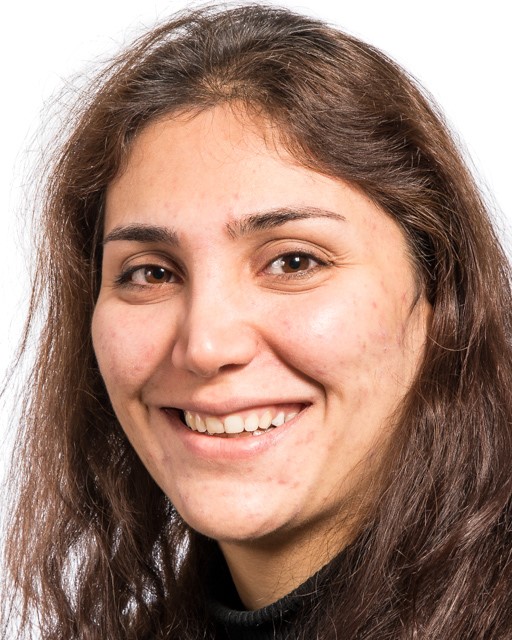}}]{Mahdieh Khanmohammadi} (M'2013), an associate professor at the University of Stavanger, has made significant contributions to the field of medical image analysis.
She was born and raised in Iran pursued her passion for science and technology by studying bioelectrical engineering as her bachelor and continued to master’s degree in the same field at the University of Tehran. 
Mahdieh obtained her second master’s degree in computational engineering from University of Halmstad in Sweden. Motivated by her master's research, Mahdieh pursued a Ph.D. program at the University of Copenhagen in Denmark. 
There, she joined the image processing group at the Computer Science Department, working on a project in collaboration with the neuroscientists at Aarhus University and the Statistical and Mathematical Science Department at Aalborg University. 

Her doctoral research involved analyzing the effect of behavioral stress on the brain, utilizing microscopic images of rat brains. Following the completion of her Ph.D., Mahdieh continued her academic journey as a postdoctoral researcher at the University of Stavanger. In collaboration with cardiologists at Stavanger University Hospital, she focused on analyzing angiographic data of the heart to estimate blood velocity.
\end{IEEEbiography}

\EOD

\end{document}